\documentclass[%
superscriptaddress,
amsmath,amssymb,
aps,
twocolumn,
]{revtex4-2}

\usepackage{hyperref}
\hypersetup{colorlinks,allcolors=blue}
\usepackage{graphicx}
\usepackage{bm}
\usepackage[all]{hypcap} 
\usepackage{color}
\usepackage{siunitx}
\usepackage{soul}
\usepackage{tabularx}

\begin{document}

\title{Universal orbital and magnetic structures in infinite-layer nickelates}

\author{M.~Rossi}
\affiliation{Stanford Institute for Materials and Energy Sciences, SLAC National Accelerator Laboratory, 2575 Sand Hill Road, Menlo Park, California 94025, USA}

\author{H.~Lu}
\affiliation{Stanford Institute for Materials and Energy Sciences, SLAC National Accelerator Laboratory, 2575 Sand Hill Road, Menlo Park, California 94025, USA}
\affiliation{Department of Physics, Stanford University, Stanford, California 94305, USA}

\author{K.~Lee}
\affiliation{Stanford Institute for Materials and Energy Sciences, SLAC National Accelerator Laboratory, 2575 Sand Hill Road, Menlo Park, California 94025, USA}
\affiliation{Department of Physics, Stanford University, Stanford, California 94305, USA}

\author{B.~H.~Goodge}
\affiliation{School of Applied and Engineering Physics, Cornell University, Ithaca, New York 14850, USA}
\affiliation{Kavli Institute at Cornell for Nanoscale Technology, Cornell University, Ithaca, New York 14850, USA}

\author{J.~Choi}
\affiliation{Diamond Light Source, Harwell Campus, Didcot OX11 0DE, United Kingdom}

\author{M.~Osada}
\affiliation{Stanford Institute for Materials and Energy Sciences, SLAC National Accelerator Laboratory, 2575 Sand Hill Road, Menlo Park, California 94025, USA}
\affiliation{Department of Materials Science and Engineering, Stanford University, Stanford, California 94305, USA}

\author{Y.~Lee}
\affiliation{Department of Physics, Stanford University, Stanford, California 94305, USA}

\author{D.~Li}
\affiliation{Stanford Institute for Materials and Energy Sciences, SLAC National Accelerator Laboratory, 2575 Sand Hill Road, Menlo Park, California 94025, USA}
\affiliation{Department of Applied Physics, Stanford University, Stanford, California 94305, USA}

\author{B.~Y.~Wang}
\affiliation{Stanford Institute for Materials and Energy Sciences, SLAC National Accelerator Laboratory, 2575 Sand Hill Road, Menlo Park, California 94025, USA}
\affiliation{Department of Physics, Stanford University, Stanford, California 94305, USA}

\author{D.~Jost}
\affiliation{Stanford Institute for Materials and Energy Sciences, SLAC National Accelerator Laboratory, 2575 Sand Hill Road, Menlo Park, California 94025, USA}

\author{S.~Agrestini}
\affiliation{Diamond Light Source, Harwell Campus, Didcot OX11 0DE, United Kingdom}

\author{M.~Garcia-Fernandez}
\affiliation{Diamond Light Source, Harwell Campus, Didcot OX11 0DE, United Kingdom}

\author{Z.~X.~Shen}
\affiliation{Stanford Institute for Materials and Energy Sciences, SLAC National Accelerator Laboratory, 2575 Sand Hill Road, Menlo Park, California 94025, USA}
\affiliation{Department of Physics, Stanford University, Stanford, California 94305, USA}
\affiliation{Geballe Laboratory for Advanced Materials, Stanford University, Stanford, California 94305, USA}

\author{Ke-Jin~Zhou}
\affiliation{Diamond Light Source, Harwell Campus, Didcot OX11 0DE, United Kingdom}

\author{E. Been}
\affiliation{Department of Physics, Stanford University, Stanford, California 94305, USA}

\author{B.~Moritz}
\affiliation{Stanford Institute for Materials and Energy Sciences, SLAC National Accelerator Laboratory, 2575 Sand Hill Road, Menlo Park, California 94025, USA}

\author{L.~F.~Kourkoutis}
\affiliation{School of Applied and Engineering Physics, Cornell University, Ithaca, New York 14850, USA}
\affiliation{Kavli Institute at Cornell for Nanoscale Technology, Cornell University, Ithaca, New York 14850, USA}

\author{T.~P.~Devereaux}
\affiliation{Stanford Institute for Materials and Energy Sciences, SLAC National Accelerator Laboratory, 2575 Sand Hill Road, Menlo Park, California 94025, USA}
\affiliation{Department of Materials Science and Engineering, Stanford University, Stanford, California 94305, USA}
\affiliation{Geballe Laboratory for Advanced Materials, Stanford University, Stanford, California 94305, USA}

\author{H.~Y.~Hwang}
\affiliation{Stanford Institute for Materials and Energy Sciences, SLAC National Accelerator Laboratory, 2575 Sand Hill Road, Menlo Park, California 94025, USA}
\affiliation{Department of Applied Physics, Stanford University, Stanford, California 94305, USA}
\affiliation{Geballe Laboratory for Advanced Materials, Stanford University, Stanford, California 94305, USA}

\author{W.~S.~Lee}
\email[Corresponding author: ]{leews@stanford.edu}
\affiliation{Stanford Institute for Materials and Energy Sciences, SLAC National Accelerator Laboratory, 2575 Sand Hill Road, Menlo Park, California 94025, USA}
\date{\today}

\begin{abstract}
We conducted a comparative study of the rare-earth infinite-layer nickelates films, RNiO$_2$ (R = La, Pr, and Nd) using resonant inelastic X-ray scattering (RIXS). We found that the gross features of the orbital configurations are essentially the same, with minor variations in the detailed hybridization. For low-energy excitations, we unambiguously confirm the presence of damped magnetic excitations in all three compounds. By fitting to a linear spin-wave theory, comparable spin exchange coupling strengths and damping coefficients are extracted, indicating a universal magnetic structure in the infinite-layer nickelates. Interestingly, while signatures of a charge order are observed in LaNiO$_2$ in the quasi-elastic region of the RIXS spectrum, it is absent in NdNiO$_2$ and PrNiO$_2$. This prompts further investigation into the universality and the origins of charge order within the infinite-layer nickelates.

\end{abstract}

\maketitle

\section{Introduction}
The discovery of superconductivity in the infinite-layer nickelates has marked a new milestone in the field of unconventional superconductivity \cite{Li2019}. While the nickelates were first thought to resemble the cuprates due to their similiar crystal structure and electron count in the 3$d$ orbitals \cite{Anisimov1999,Lee2004}, the initial stages of the experimental investigations have revealed notable differences between the two systems \cite{Li2020,Zeng2020,Lee2023,Hepting2018,Rossi2021,Goodge2021,Rossi2022}. As a result, the nickelates have emerged as a new class of strongly correlated materials, presenting fresh opportunities to study the role of strong electronic correlations in unconventional superconductors. 

To date, superconductivity has been found in La-, Pr-, and Nd-based infinite-layer nickelate thin films \cite{Li2020, Osada2020, Osada2021}. The doping-temperature phase diagram in these compound appears to be quite similar, indicating a similar underlying microscopic origin. However, subtle differences have also been observed. For example, London penetration depth measurements indicate that La- and Pr-NiO$_2$ exhibit a node in the superconducting order parameter, whereas it is less clear in NdNiO$_2$, possibly due to the magnetic moment carried by the Nd ions \cite{harvey2022}. Another example is the recent demonstration of substantial variations in the anisotropy of the upper critical field across La, Pr, and Nd-NiO$_2$ \cite{Wang2023}. These observations imply that the rare-earth element may play a role in the low-energy electronic behavior.

Indeed, theories have suggested that 5$d$ orbitals of the rare-earth element hybridize with the Ni 3$d$ orbitals and gives rise hole-like Fermi surface (FS) pockets, making the infinite-layer nickelates multi-orbital systems \cite{Nomura2019,Adhikary2020,Botana2020,Gu2020a,Kapeghian2020,Lechermann2020,Leonov2020,Liu2020,Sakakibara2020,Wu2020,Been2021}. The minimal model should at least consist of a hole-like band with dominant Ni 3$d_{x^2-y^2}$ character which is coupled to an electron-like band with a dominant character of rare-earth 5$d$ orbitals. It has been an intriguing question regarding whether rare earth element could serve as a tuning knob to control the properties of the infinite-layer nickelates, as it does in the well-known cases of the perovskite nickelates \cite{Torrance1992,Middey2016}. Along this line, theoretical studies have investigated the variation of electronic, magnetic, and lattice structures as a function of rare earth elements \cite{Been2021,Zhang2023,Subedi2023}, suggesting the possible occurrence of nontrivial phase transitions. Thus, it is of great interest to investigate the rare-earth dependence by examining the microscopic behaviors via spectroscopic measurements.

Owing to the element-specific capability, resonant inelastic x-ray scattering (RIXS) has been a powerful tool to reveal the electronic structure \cite{Hepting2018}, magnetic excitations \cite{Lu2021,gao2022}, and charge order in the infinite-layer nickelates \cite{Rossi2022,Tam2022,Krieger2022,Ren2023}. To date, most of the RIXS studies were focused on one family of the infinite-layer nickelates from various sample sources, which is not ideal for a systematic study between La-, Pr-, and Nd-NiO$_2$. In this article, we report RIXS data that were measured with identical measurement conditions on Nd-, Pr-, and La-NiO$_2$ thin films. Importantly, the samples were grown and prepared using the same synthesis and characterization standard. Since uncapped infinite-layer nickelates are prone to heterogeneity, and their synthesis is hard to control \cite{Lee2020}, we focus our studies on samples with a protective SrTiO$_3$ capping layer, for which we have previously demonstrated uniform crystallinity throughout the the thickness of the nickelate film \cite{Osada2021,Lee2020,Osada2020b}. We show that the gross features of orbital configurations are essentially identical with some minor variations in the detailed hybridization. For the low-energy excitations, we compare the dispersion and the bandwidth of the magnetic excitations in the three compounds and conclude that the magnetic structure is universal in the nickelates. Interestingly, in the quasi-elastic RIXS spectra, we found signatures of a charge order only in LaNiO$_2$, but not in NdNiO$_2$ and PrNiO$_2$. In the context of recent debates about the charge order in infinite-layer nickelates \cite{raji2023, parzyck2023}, we leverage cross-sectional scanning transmission electron microscopy (STEM) to investigate the secondary phases present in our LaNiO$_2$ film, and discuss the implications on the observed charge order peak in the LaNiO$_2$ film.

\begin{figure}
	\centering
	\includegraphics[width=\columnwidth]{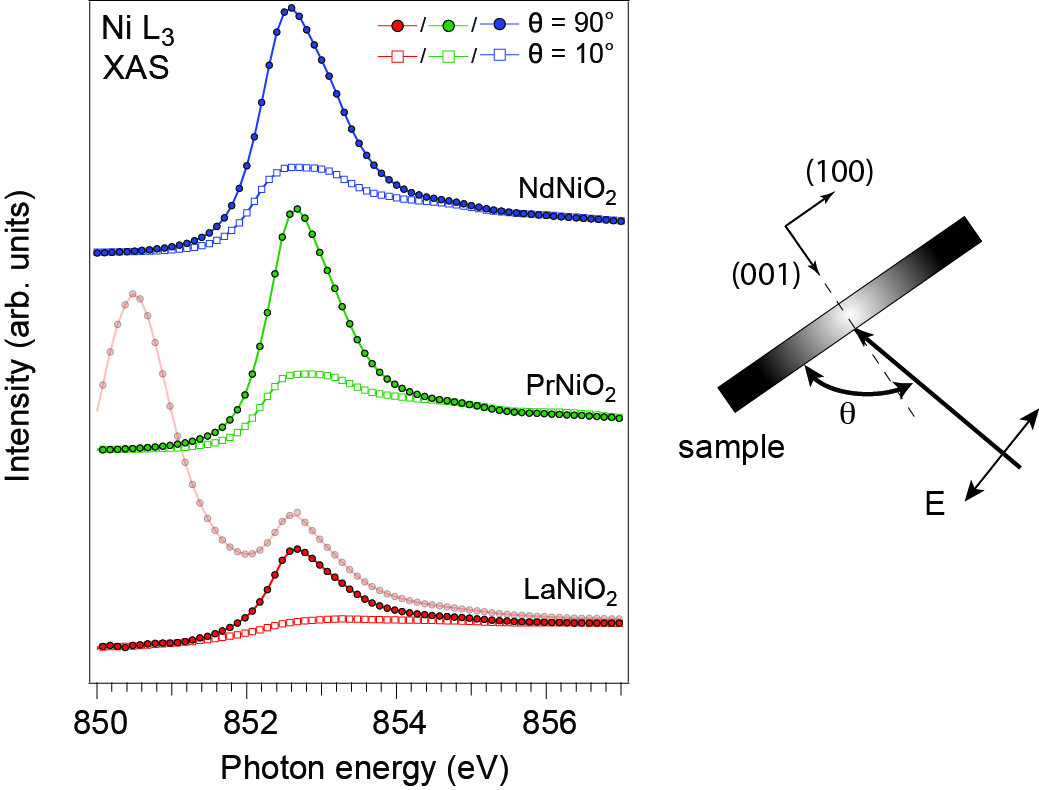}
	\caption{\label{XAS} Ni L$_3$-edge XAS of NdNiO$_2$, PrNiO$_2$, and LaNiO$_2$ obtained by total electron yield in both normal (solid circles, $\theta = 90^{\circ}$) and grazing incident (open squares, $\theta=10 ^{\circ}$) geometry. The measurement geometry is sketched in the right panel. A constant background has been removed and the post-edge intensity has been normalized for comparing XAS taken at both geometries. The raw XAS of LaNiO$_2$ (light red) is superimposed with the the La-M$_4$-edge-removed XAS (dark red) for reference. The data of NdNiO$_2$ is adapted from Ref. \cite{Rossi2021}.}
\end{figure}

\begin{figure*}
	\centering
	\includegraphics[width=\textwidth]{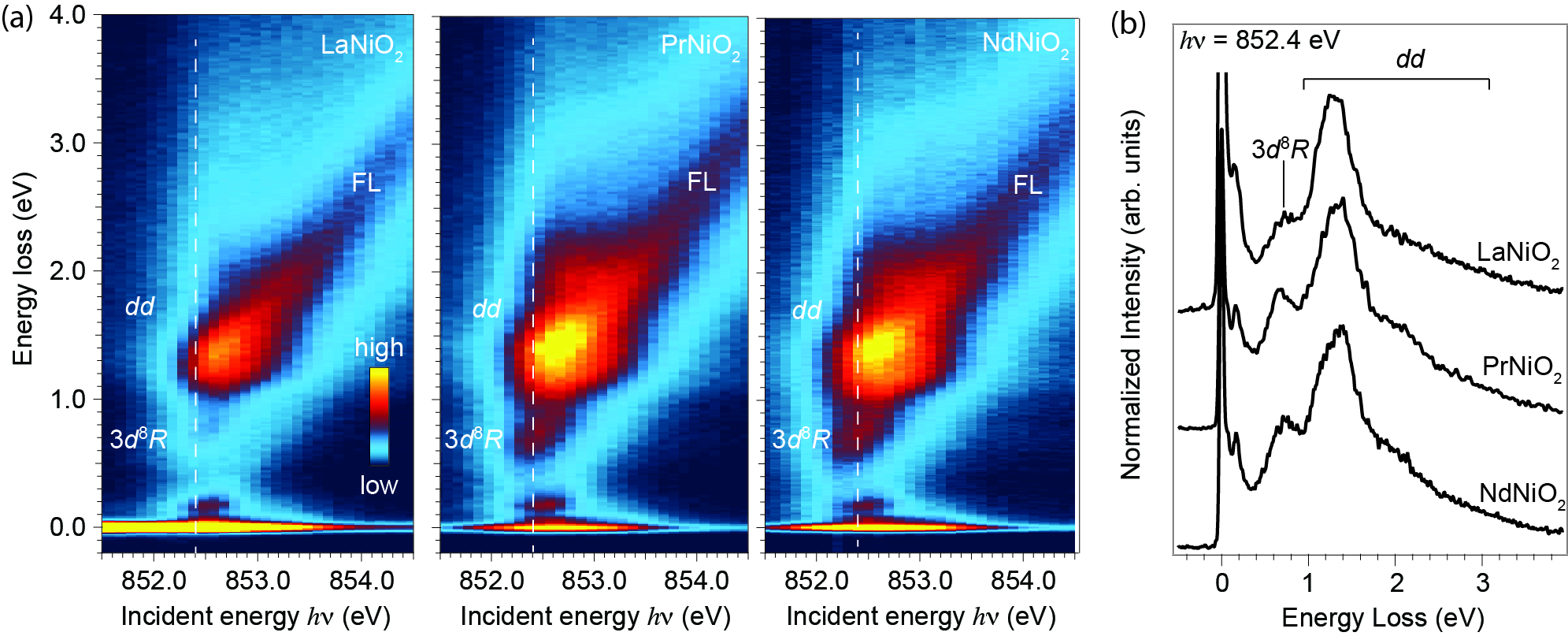}
	\caption{\label{RIXS_Map} (a) RIXS intensity maps of LaNiO$_2$, NdNiO$_2$, and PrNiO$_2$ as a function of incident photon energy across the Ni $L_3$ edge. 3$d^8R$ denotes the spectral feature at approximately 0.65 eV. $dd$ and FL indicate the excitations within the Ni 3$d$ orbitals and the fluorescence emission, respectively. (b) The energy loss spectra taken at the incident photon energy of 852.4 eV, as indicated by the white dashed lines in (a). Data of NdNiO$_2$ are adapted from Ref.~\onlinecite{Lu2021}.}
\end{figure*}

\section*{Experiment details}
Thin films of the precursor perovskite $R$NiO$_3$ ($R$ = La, Nd, Pr) with a thickness of 10 nm were grown on a substrate of SrTiO$_3$(001).  The $c$-axis oriented infinite-layer $R$NiO$_2$ was obtained by employing a topotactic reduction process \cite{Lee2020}. To protect and support the crystalline order, a capping layer made of five unit cells of SrTiO$_3$(001) was grown on top of the nickelate films before the topotactic reduction. X-ray diffraction (XRD) measurements were conducted to confirm the quality of the infinite-layer nickelate. The XRD data were included in the Supplementary Material \cite{SM}. XAS and RIXS measurements were performed at beamline I21 of the Diamond Light Source (United Kingdom). The combined energy resolution of the RIXS measurements was approximately 40 meV at the Ni $L_3$ edge. For the incident energy RIXS map (Fig. \ref{RIXS_Map}), the spectra were collected at an incidence angle of \SI{35}{\degree} and scattering angle of \SI{154}{\degree}. Measurements were taken at a temperature of 20 K. For the $dd$ excitations, magnetic excitations, and quasi-elastic map, the RIXS spectra were taken at the photon energy of 852.6 eV at which the intensity of the $dd$ and magnetic excitations is maximal. Since the magnetic excitations are quasi-two-dimensional \cite{Lu2021}, the momentum dependent data have been denoted as a function of the projected in-plane momentum transfer $q_{\parallel}$, which can be varied by rotating the sample angle with a fixed scattering angle. For all the momentum-dependent RIXS data shown here were obtained with the scattering angle set to 154$^\circ$ using $\pi$ polarization of the incident photons with grazing exit geometry, at which the magnetic cross section is dominant.

Cross-sectional specimens for STEM analysis were prepared by the standard focused ion beam (FIB) lift-out process on a Thermo Fisher Helios G4 UX FIB and imaged on an aberration-corrected FEI Titan Themis 300 operating at 300 kV with a probe convergence angle of 21.4 mrad and inner (outer) collection angles of 68 (200) mrad.

\section*{results}
We begin by examining the orbital configurations of La-, Nd-, and Pr-NiO$_2$. Figure \ref{XAS}~(a) illustrates the x-ray absorption spectra (XAS) near the Ni $L_3$-edge for these three parent compounds. In the normal incident geometry ($\theta = 90^\circ$), where the polarization of the incident photons aligns with the Ni-O bond direction, all XAS exhibit a single main peak, as reported in previous experiments \cite{Hepting2018, Lu2021, gao2022, Rossi2021, Krieger2022, Ren2023}. We note that the absorption is significantly reduced in the grazing incident geometry ($\theta = 10^\circ$), where the majority of the incident photon polarization lies in the normal direction of the NiO$_2$ plane. This observation indicates a significant anisotropy between the in-plane and out-of-plane orbital configurations of the system. The presence of a single XAS peak and the linear dichroism in XAS support the notion of a quasi-two dimensional electronic structure, primarily consisting of a dominant 3$d^9$ character with a half-filled 3$d_{x^2-y^2}$ orbital \cite{Hepting2018,Rossi2021,Goodge2021}.

Further insights into the orbital configuration can be obtained from RIXS spectra across the Ni $L_3$-edge. As depicted in Fig. \ref{RIXS_Map} (b), the overall features of the RIXS incident map exhibit similarities among the three compounds. These include magnetic excitations below an energy of 0.2 eV, a spectral feature around $\sim$0.65 eV (referred to as 3$d^8R$), $dd$ excitations within the 3$d$ orbitals in the energy range of 1.0 to 2.5 eV, and fluorescence emission (FL) above 2.5 eV. The presence of similar energy scales for $dd$ and FL excitations reflects a generic crystal field splitting characteristic of the three families of nickelates, as expected. Notably, the 3$d^8R$ feature has been associated with an excitation related to the hybridized rare-earth 5$d$ ligands and Ni 3$d$ states \cite{Hepting2018}. Its observation in all three compounds indicates that it is a common feature of infinite-layer nickelates. This observation aligns with various theoretical predictions that suggest a multi-orbital nature in infinite-layer nickelates, where the rare-earth 5$d$ states contribute to the band structures near the Fermi energy \cite{Nomura2019, Adhikary2020, Botana2020, Gu2020a, Kapeghian2020, Lechermann2020, Leonov2020, Liu2020, Sakakibara2020, Wu2020, Been2021}. This is in contrast to high-$T_\mathrm{c}$ cuprates, where the states near the Fermi energy are primarily dominated by a mixture of Cu 3$d_{x^2-y^2}$ character and the O 2$p$ ligand states. 

\begin{figure}
	\centering
	\includegraphics[width=\columnwidth]{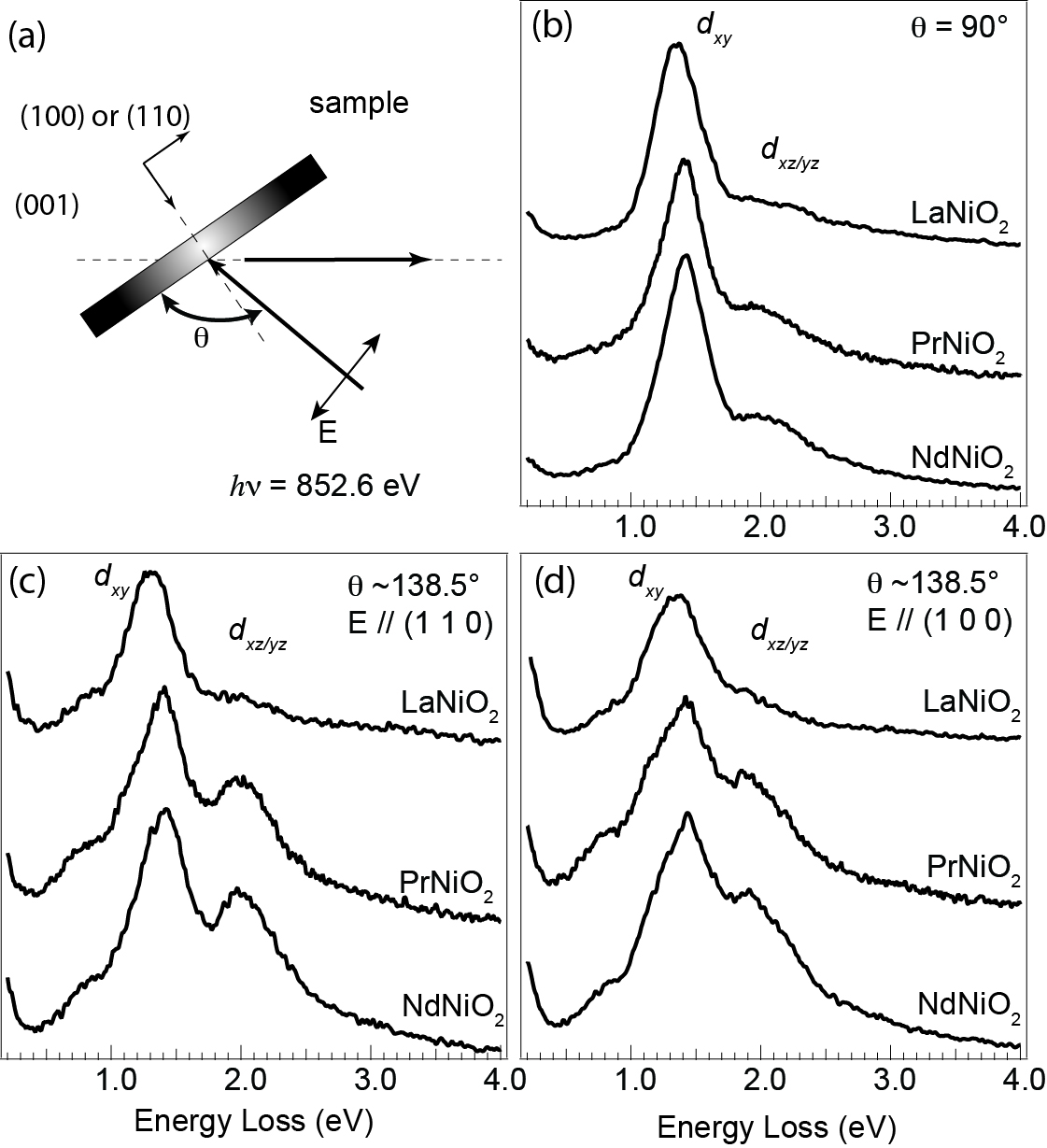}
	\caption{\label{ddExcitation} (a) A sketch of the $\pi$-scattering geometry used for taking the data reported in this article. In this geometry, the electric field polarization of the incident photon (denoted as "E") is in the scattering plane. RIXS spectra of LaNiO$_2$, NdNiO$_2$, and PrNiO$_2$ at (b) $\theta = 90^{\circ}$ and (c) $\theta \sim 138.5^{\circ}$ with the (100) direction (\textit{i.e.} the Ni-O bond direction) in the scattering plane. (d) RIXS spectrum taken at $\theta \sim 138.5^{\circ}$ with (110) direction in the scattering plane.}
\end{figure}

Despite the overall similarity, we have observed a subtle difference in the $dd$ excitations. Figure \ref{ddExcitation} illustrates the $dd$ excitations obtained at three different angles, effectively varying the incident photon polarization with respect to the crystallographic axes. As a result, the cross-section of $dd$ excitations associated with different orbital symmetries are modulated \cite{Moretti2011}. Specifically, in the normal incident geometry ($\theta = 90^\circ$) depicted in Figure \ref{ddExcitation}(a), the electric field of the incident photon polarization lies in the NiO$_2$ plane, favoring transitions associated with in-plane orbitals such as $d_{x^2-y^2}$ and $d_{xy}$. In the other geometry ($\theta \sim 138.5^\circ$ in Figure \ref{ddExcitation}), the photon polarization now includes a notable component perpendicular to the NiO$_2$ plane, enabling excitations associated with out-of-plane $d_{xz/yz}$ orbitals. Since the $dd$ excitations in the RIXS incident photon map are generic (Fig. \ref{RIXS_Map} (a)), we follow the assignment of $dd$ peaks described in our previous work \cite{Rossi2021}, as also denoted in Figure \ref{ddExcitation}. Interestingly, as shown in Figure \ref{ddExcitation}(c) and (d), while the excitations associated with the $d_{xz/yz}$ orbitals are prominent in Nd- and Pr-NiO$_2$, they are significantly weaker in LaNiO$_2$. This observation suggests that LaNiO$_2$ exhibits more extended $d_{xz/yz}$ orbitals. 


\begin{figure*}
	\centering
	\includegraphics[width=\textwidth]{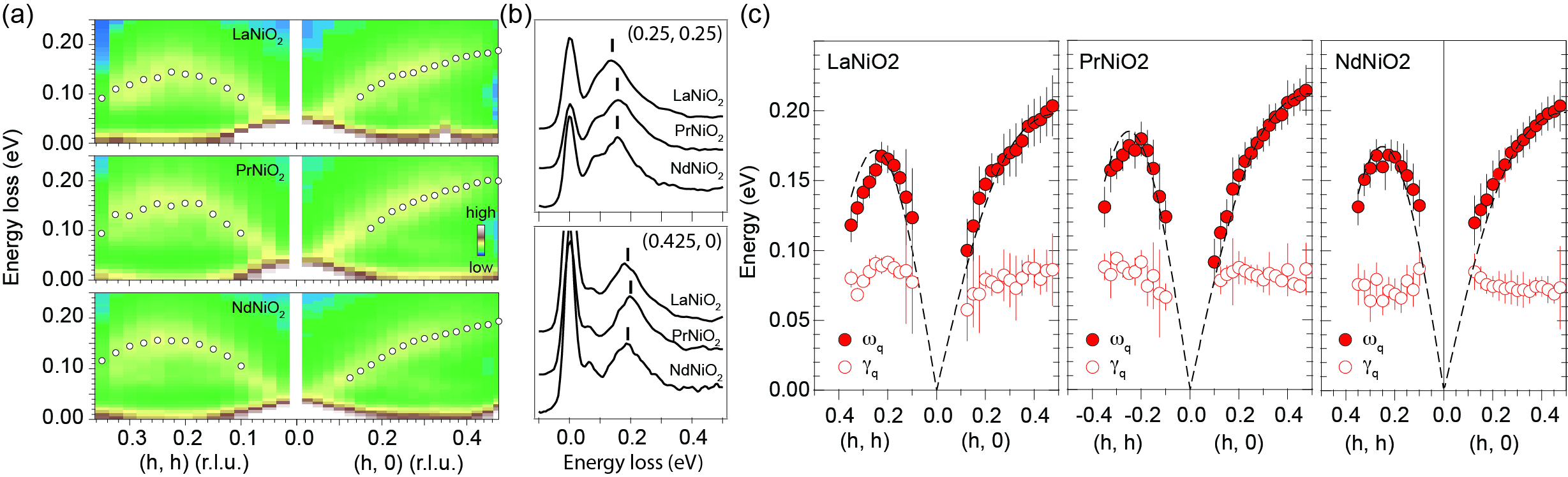}
	\caption{\label{Magnetic} (a) RIXS intensity map of LaNiO$_2$, NdNiO$_2$, and PrNiO$_2$, as a function of in-plane momentum along (h, 0) and (h,h) directions. The white circles indicate peak positions of the magnetic excitations. (b) RIXS spectra of the three nickelates taken at two representative momenta near the maxima of the dispersion. The ticks are guides-to-the-eye for the peak position of the magnetic excitation spectra. (c) The extracted magnetic mode energy $\omega_q$ and damping coefficients $\gamma_q$ from fitting the data to the damped harmonic oscillation function for the three compounds. The linear spin wave fits (dashed lines) to the extracted dispersions are also superimposed. The data of NdNiO$_2$ is adapted from Ref. \cite{Lu2021}. The raw data of La- and PrNiO$_2$ are included in the Supplementary Material \cite{SM}.}
\end{figure*}

\begin{table}[b!]
\centering
\begin{tabularx}{0.9\columnwidth}{| >{\raggedright\arraybackslash}X | >{\centering\arraybackslash}X | >{\raggedleft\arraybackslash}X |>
{\raggedleft\arraybackslash}X | } 
 \hline
  & $J_1$ (meV)& $J_2$ (meV)& \textit{c}-axis (\AA) \\ 
 \hline
 LaNiO$_2$ & 61.6 $\pm$ 4.2 & -11.3 $\pm$ 3.0 & 3.39 \\ 
 PrNiO$_2$ & 66.9 $\pm$ 3.2 & -11.4 $\pm$ 2.3 & 3.31 \\
 NdNiO$_2$ & 64.1 $\pm$ 3.4 & -10.2 $\pm$ 2.4 & 3.28 \\
\hline
\end{tabularx}
\caption{A summary of the nearest neighbor and next nearest neighbor spin exchange coupling obtained by fitting the dispersion to a linear spin wave theory.}
\label{table:magnetic}
\end{table}

Next, we discuss the magnetic excitations, which bear the hallmarks of the underlying electronic correlations \cite{Lu2021}. Figure \ref{Magnetic}(a) shows the unambiguous presence of a branch of dispersive magnetic excitations in all three compounds with similar characteristics. Specifically, the dispersion emanates away from the zone center and reaches respective maxima near (0.5, 0) and (0.25, 0.25), consistent with the spin wave dispersion in an antiferromagnetically coupled spin-1/2 square lattice system. By comparing the RIXS spectra near the two maxima, as shown in Figure \ref{Magnetic}(b), we found that the peak positions of the magnetic excitations are essentially identical in the three compounds within the uncertainty of our experiments. This suggests that the bandwidth of the magnetic excitations are quite similar across the three families of nickelates. Additionally, we observe that the magnetic peak spectra are broad (Figure \ref{Magnetic}(b)), indicating that the excitations are heavily damped. This is consistent with the presence of rare-earth 5$d$ FS pocket in the undoped parent compounds, which can dissipate the magnetic excitations in the particle-hole continuum \cite{Lu2021}.

To obtain a more quantitative comparison, following our previous work on NdNiO$_2$ \cite{Lu2021}, we fit the RIXS magnetic spectrum to a damped harmonic oscillation functions $\chi''(q,\omega)$ \cite{Lamsal2016}, given by 
\begin{equation}
\chi''(q,\omega) = \frac{\gamma_q \omega}{(\omega^2-\omega_q^2)^2 + 4\gamma_q^2\omega^2}
\end{equation}
where $\omega_q$ is the undamped mode energy and $\gamma_q$ damping coefficient. The extracted $\omega_q$ and $\gamma_q$ are summarized in Fig. \ref{Magnetic}(c). All three nickelates exhibit a similar energy-momentum dispersion with a similar degree of damping. Then, we fit the extracted dispersion to a linear spin wave form for the spin-1/2 square-lattice Heisenberg antiferromagnet \cite{Coldea2001}, including nearest- and next-nearest-neighbor exchange couplings,

\begin{equation}
H = J_1 \sum_{i,j}S_i \cdot S_j + J_2 \sum_{i,i'}S_i \cdot S_{i'}
\end{equation}
where $S_i$, $S_j$ and $S_{i'}$ denote Heisenberg spins at site $i$, nearest-neighbor sites $j$, and the next- 
nearest-neighbor sites $i'$, respectively. The fitted $J_1$ and $J_2$ are summarized in Table \ref{table:magnetic}. Both $J_1$ and $J_2$ are comparable in all three nickelates, except that the $J_1$ of PrNiO$_2$ appears to be slightly larger than the others, consistent with the higher peak position in the raw magnetic spectra shown in Fig. \ref{Magnetic}(b). This finding aligns with the weak rare-earth dependence on the spin exchange interaction predicted by theoretical models \cite{Been2021, Zhang2023}. We also note that there appears to be no clear correlation with the \textit{c}-axis lattice parameter, which monotonically decreases by replacing the rare-earth element from La to Nd. This null correlation suggests that the spin exchange interaction is essentially insensitive to the electronic structure along the \textit{c}-axis direction. Thus, the magnetic interactions are dictated by the in-plane low energy electronic structure, presumably the 3d$_{x^2-y^2}$ orbitals.

It is important to emphasize that our results do not necessarily imply the existence or nonexistence of long-range antiferromagnetic order (AFM), as the putative AFM wave-vector is beyond the reach of the Ni $L$-edge RIXS. Indeed, multiple studies of bulk poloycrystalline sample do not shown long range AFM order \cite{Hayward2003, Lin_2022,Ortiz2022}.  Nevertheless, the observation firmly establishes the universal presence of substantial antiferromagnetic spin interactions in infinite-layer nickelates, suggesting that Mott physics should play a significant role in shaping the microscopic electronic structure of these nickelates.

\begin{figure}
	\centering
	\includegraphics[width=\columnwidth]{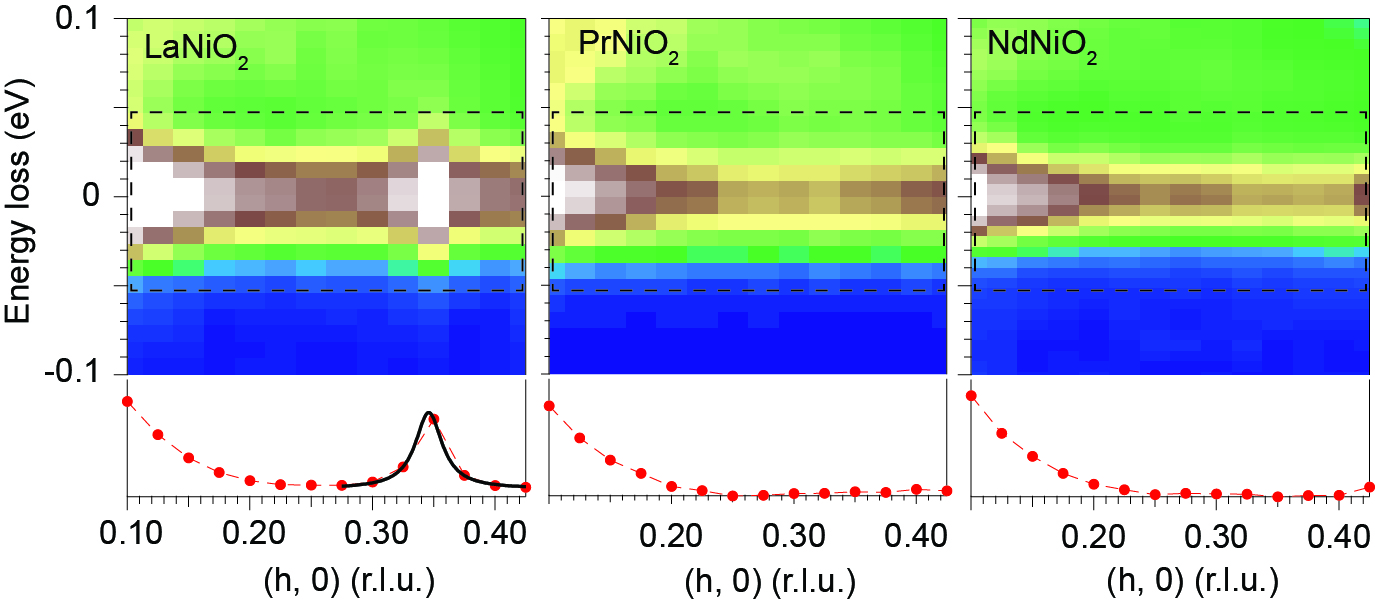}
	\caption{\label{quasielastic} (a) Quasi-elastic RIXS intensity map of LaNiO$_2$, PrNiO$_2$, and NdNiO$_2$, as a function of in-plane momentum along the (h,0) directions. Momentum distribution of the integrated intensity within the black dashed box is shown in lower panels as red circles connected by dashed lines. The solid black curve is a fit to a Lorentzian function for the charge order peak with a fitted peak position of 0.345$\pm$0.0007 r.l.u..}
\end{figure}

Another hallmark of strong electronic correlations is a complex phase diagram comprising multiple quantum phases. Particularly, those quantum phases that break the translation symmetry can be detected by examining the momentum distribution of the quasi-elastic peak intensity in the RIXS spectrum (See, for example, Ref. \onlinecite{Ghiringhelli2012YBCO}). Indeed, previously, we observed a charge order in LaNiO$_2$ and investigated its doping and temperature dependence using RIXS in $\sigma$-scattering polarization \cite{Rossi2022}, where the incident x-ray polarization is perpendicular to the scattering plane. Using a different scattering geometry (\textit{e.g.}, $\pi$-polarization), as shown in Fig. \ref{quasielastic}, a peak at (0.345$\pm$0.0007, 0) is observed in LaNiO$_2$, confirming the presence of the charge order. Interestingly, the signature of the charge order scattering peak is absent in the RIXS data of Nd- and Pr-NiO$_2$. While this may appear to indicate that the charge order is not universal in infinite-layer nickelates, recent findings have presented a puzzle. Specifically, a peak with a wavevector of (0.333, 0) has been reported in NdNiO$_2$ films prepared without a SrTiO$_3$ capping layer \cite{Tam2022,Krieger2022} and PrNiO$_2$ with a STO capping layer \cite{Ren2023}. Notably, the peak intensity in uncapped NdNiO$_2$ exhibits a positive correlation with the spectral weight of the RIXS 3$d^8R$ feature \cite{Krieger2022,Tam2022}, but our LaNiO$_2$ exhibits the least pronounced 3$d^8R$ peak in RIXS spectra of the three families (Fig. \ref{RIXS_Map}(b)). Consequently, the emergence of CO appears to be sample-specific. Some aspects of the material specificity will be further discussed in the next section. 

\begin{figure*}
	\centering
	\includegraphics[width=\textwidth]{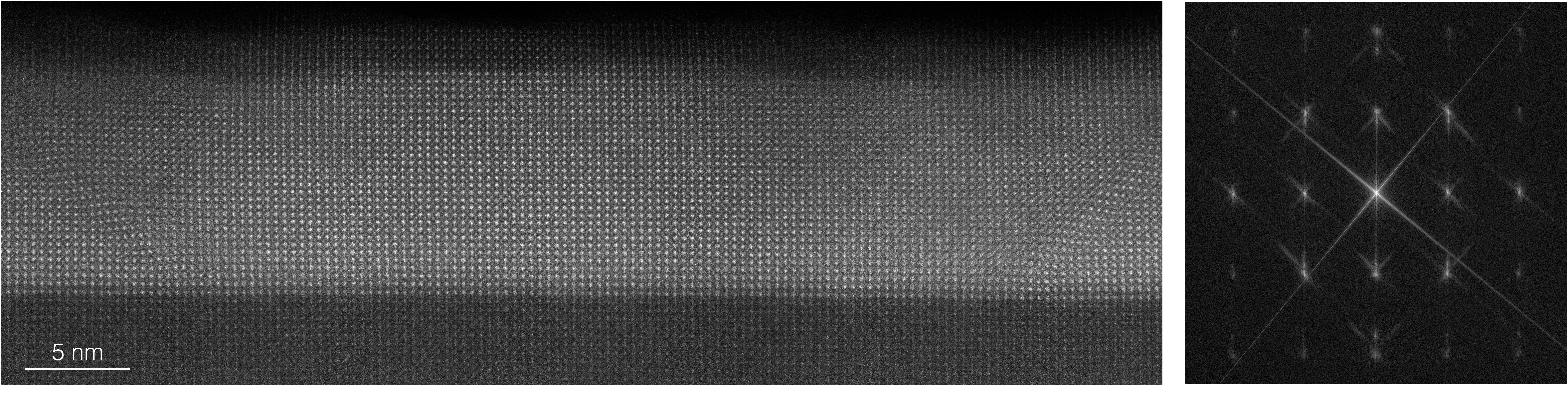}
	\caption{\label{HAADF_STEM_large} (left) A representative HAADF-STEM image and (right) the corresponding fast Fourier transform (FFT) of a large area in a LaNiO$_2$ thin film on SrTiO$_3$.}
\end{figure*}
\begin{figure*}
	\centering
	\includegraphics[width=\textwidth]{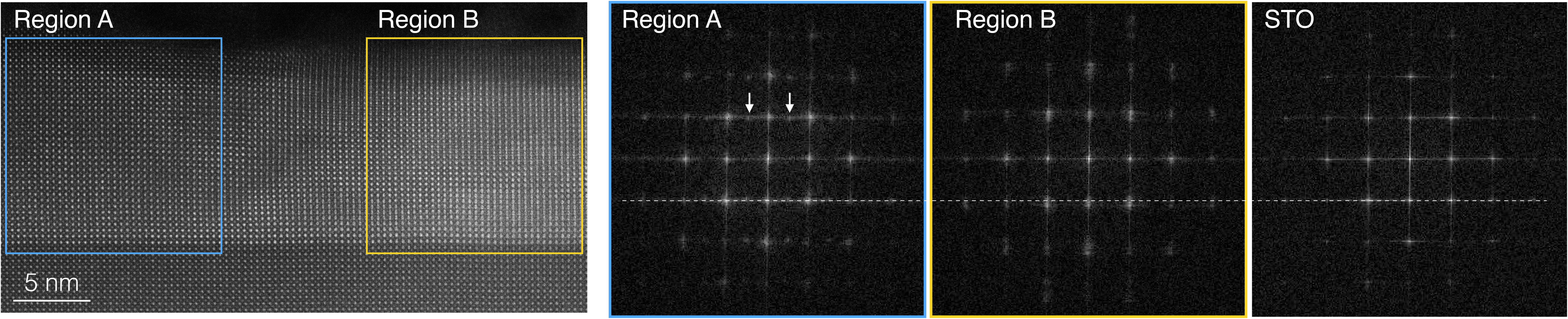}
	\caption{\label{HAADF_STEM_mix_oxygen} HAADF-STEM image and selected area FFTs of a region in the same LaNiO$_2$ film which shows mixed oxygen stoichiometry. The FFTs of Region A (blue box, un-reduced perovskite nickelate) and B (infinite-layer nickelate) are shown in right panels; the far-right FFT (STO) is selected from a region of the SrTiO$_3$ substrate in the same image. The dotted line marks the out-of-plane peak distance in the SrTiO$_3$ substrate, corresponding to a lattice spacing of 3.905 \AA, indicating a reduced \textit{c}-axis lattice constant in Region B.}
\end{figure*}

\begin{figure}
	\centering
	\includegraphics[width=\columnwidth]{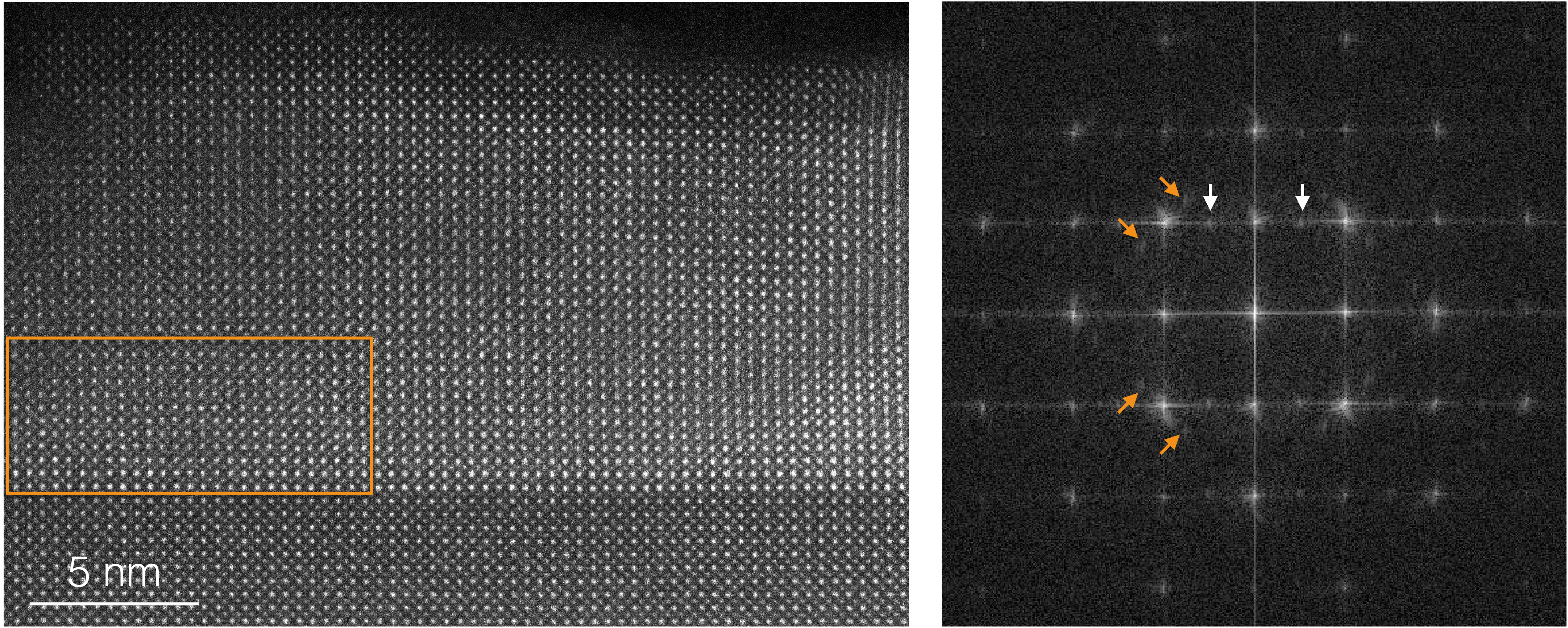}
	\caption{\label{HAADF_zoom_in} Higher-magnification HAADF-STEM image of Region A in Fig.\ref{HAADF_STEM_mix_oxygen} and the associated FFT. No splitting in the \textit{c}-axis peaks is observed, indicating the nominally perovskite phase of this region. Half-order in-plane peaks (white arrows) are clearly visible as discussed in Fig. \ref{HAADF_STEM_mix_oxygen}, as well as faint (1/4, 0, 1/4) peaks in addition to the primary peaks (orange arrows) in the region marked by the orange box.}
\end{figure}

\section*{Discussion} 
As previously discussed by Krieger \textit{et al.} \cite{Krieger2022}, the XAS and RIXS spectra of NdNiO$_2$ without a SrTiO$_3$ capping layer exhibit distinct characteristics compared to their counterparts with a capping layer. Namely, the XAS exhibits less linear dichroism between $\sigma$ and $\pi$ polarization of the incident photon, indicating a more three dimensional orbital configuration in the uncapped NdNiO$_2$. In addition, the 3$d^8R$ feature in the RIXS spectra also becomes stronger and resonates at a photon energy lower than the main peak in Ni $L$-edge XAS, producing a shoulder in the leading edge of the XAS \cite{Hepting2018, Krieger2022, Tam2022}. Our results shown in Fig. \ref{XAS} and \ref{RIXS_Map}  unambiguously establish the generic orbital configuration in capped La-, Pr-, and Nd-NiO$_2$ films and confirm the different electronic configurations between the capped and uncapped nickelate films. This notion gained further support from the work by Raji \textit{et al.} via STEM-EELS and hard x-ray photoemission spectroscopy \cite{raji2023}. We also note that in the uncapped NdNiO$_2$ thin film, no magnetic excitations were resolved, raising a question regarding the dichotomy between the charge order and the spin order that could be switched by the presence of the capping layer \cite{Krieger2022}. We remark that this needs not be the case, as magnetic excitations and charge order coexist in our capped LaNiO$_2$ (Fig. \ref{Magnetic}).

It is important to note that the recent discovery of charge order in infinite-layer nickelates has sparked a significant debate. Pelliciari \textit{et al.} proposed that this observation could have arisen from the SrTiO$_3$ (STO) substrate's (1 0 1) Bragg peak due to the third harmonic x-ray photon contamination originating from the beamline \cite{pelliciari2023comment}. Since all of our films were grown on STO substrates and capped with STO layers, we would have expected to observe the purported Bragg peak contamination in all of our samples. However, this is contradicted by the data presented in Fig. \ref{quasielastic}. In addition, we remark that Tam \textit{et al} have conducted comprehensive measurements to specifically address Pelliciari's scenario \cite{tam2023reply}. 

Another scenario may be more difficult to completely rule out. Some recent studies propose that the "charge order" peak in Nd-based compounds originates from secondary phases, where ordered rows of excess apical oxygens with 3 unit cell periodicity (denoted as 3$a_o$) are formed \cite{parzyck2023, raji2023}. In fully reduced NdNiO$_2$, no charge order is detected, as found in our NdNiO$_2$ and PrNiO$_2$ samples (Fig. \ref{quasielastic}). But are secondary phases also present in our LaNiO$_2$ sample?

We first remark that while we cannot comprehensively rule out the presence of secondary phases, our LaNiO$_2$ film should predominantly consist of the infinite-layer nickelate phase, with a purity comparable to that of the NdNiO$_2$ and PrNiO$_2$ samples. This assertion is supported by the consistent spectroscopic features observed in XAS (Fig. \ref{XAS}), RIXS maps (Fig. \ref{RIXS_Map}), and RIXS magnetic excitation spectra (Fig. \ref{Magnetic}) across the three families of nickelates. Also, the charge order in LaNiO$_2$ has an incommensurate wave-vector, unlike the commensurate 3$a_o$ unit-cell periodicity found in the uncapped NdNiO$_2$ and the secondary phase \cite{parzyck2023, raji2023}. Moreover, our previous work demonstrates a systematic doping evolution of the charge order wavevctor, width, and onset temperature, inconsistent with an explanation solely based on oxygen order of secondary phases \cite{Rossi2022}.

In order to gain further insight about secondary phases in our LaNiO$_2$ films, we conduct a detailed structural investigation of a capped LaNiO$_2$ thin film grown on SrTiO$_3$ with identical conditions to that studied by RIXS using high-angle annular dark-field (HAADF) scanning transmission electron microscopy (STEM). Figure \ref{HAADF_STEM_large} shows a representative HAADF-STEM image of a large area of the LaNiO$_2$ thin film and the fast Fourier transform (FFT) of the micrograph which provides quantitative analysis of the periodicities present in the image. The diagonal canting which is observed on either side of the HAADF-STEM image in Fig. \ref{HAADF_STEM_large} has been discussed elsewhere \cite{Osada2021} and is attributed to a local reorientation of the \textit{c}-axis into the plane of the film as a mechanism for relieving the relatively high compressive strain imposed on the reduced LaNiO$_2$ by epitaxial growth on SrTiO$_3$. 

As sampled by our HAADF-STEM measurements, most of the film is consistent with the reduced infinite-layer structure and certain types of heterogeneity in the lattice (e.g., Ruddlesden-Popper type stacking faults) which have been well documented throughout all reported infinite-layer nickelate thin film species \cite{Lee2020, Li2020, Osada2020b, Osada2021,Zeng2022}. We do, however, observe some limited regions of off-stoichiometry from the infinite-layer formula LaNiO$_2$. But rather than the 3$a_o$ order in the excess oxygen, we find instead local regions of film which appear consistent with the nominally unreduced perovskite precursor phase, i.e., LaNiO$_3$. The HAADF-STEM image in Fig. \ref{HAADF_STEM_mix_oxygen} shows one region with local perovskite- and infinite-layer like regions on either side of the $\sim$40 nm field of view, separated by a diagonal canted region in the middle of the image. The right-hand region marked by a yellow box (Region B) and its corresponding FFT show a lattice which is consistent with the infinite-layer phase: namely a reduced \textit{c}-axis lattice constant and lack of additional superlattice peaks. The left-hand region marked by the blue box (Region A) and its corresponding selected area FFT show a lattice structure which is consistent with the perovskite precursor phase, characterized in particular by the half-order in-plane (1/2, 0, 0) superlattice peaks and the expanded \textit{c}-axis lattice constant similar to that of the SrTiO$_3$ substrate (STO in Fig. \ref{HAADF_STEM_mix_oxygen}). A subtle (1/4, 0, 1/4) ordering can be found in the few unit cells nearest to the substrate-film interface within this perovskite-like region, as marked by the orange box and arrows the HAADF-STEM image and corresponding FFT shown in Figure \ref{HAADF_zoom_in}. 

Across a set of $\sim$40 images constituting a quasi-random survey spanning $\sim$1.5 $\mu$m in this LaNiO$_2$ thin film, we find no signs of the (1/3, 0, 1/3) peaks of the 3$a_o$ order described in similar studies of NdNiO$_2$ thin films. It is important to note that this lack of evidence for 3$a_o$ oxygen ordering does not definitively rule out the possibility of such ordering elsewhere in the film, but rather provides an estimated upper bound on the prevalence any such ordering we would expect in this and similar samples. Still, it is further interesting to reflect that the 3$a_o$ ordering which has been observed by similar techniques in uncapped NdNiO$_2$ thin films reflects a different character of accommodating oxygen off-stoichiometry than what we observe here \cite{raji2023}. Here, we find nm-scale regions which appear more or less fully perovskite-like interspersed between large regions of what appear to be fully reduced infinite-layer, while reports in the uncapped NdNiO$_2$ thin films suggest a wider or more disperse distribution of excess oxygen into regions which can be locally considered NdNiO$_{2+\delta}$ ($\delta$ = 0.33 - 0.66) \cite{raji2023,parzyck2023}. We speculate that these differences in the oxygen distribution could be related to different intrinsic effects of the La- and Nd-based compounds as well as to more extrinsic consideration such as lattice defects which may act as barriers or channels for oxygen de-intercalation during the chemical reduction process. A more thorough and comprehensive understanding of these differences is called for, and should include comparison of samples with different rare earth cations (La, Nd, and even Pr) synthesized by different research groups. 

Finally, some characteristics of LaNiO$_2$ are noteworthy. As suggested by Been \textit{et al.} \cite{Been2021}, due to the presence of empty 4$f$ states, the calculated size of the La 5$d$ Fermi surface (FS) pocket and the effective mass along the out-of-plane direction are the smallest among all rare-earth infinite layer nickelates and do not follow the same trend as others. Additionally, theory has also suggested that the modulation of charge order might be intimately related to both the rare-earth FS and the Ni 3$d_{x^2-y^2}$ states \cite{peng2022}. Therefore, it is possible that the La 5$d$ FS in LaNiO$_2$ is somehow more conducive to the emergence of charge order. Another characteristic of LaNiO$_2$ is its epitaxial strain. By comparing the lattice constants to those of the bulk crystal powders \cite{Lin_2022}, we determined that the La-, Pr-, and Nd-NiO$_2$ thin films exhibit compressive epitaxial strains of 1.34\%, 0.9\%, and 0.6\%, respectively. Interestingly, the LaNiO$_2$ film experiences the most compressive strain compared to the other two families of nickelates. Drawing from the insights gained from the study of cuprates, where the emergence and behavior of charge order can be substantially influenced by strain \cite{Kim2018,Bluschke2018j, Choi2022, Wang2022, gupta2023}, it may be reasonable to speculate whether strain plays a role in the emergence of charge order in infinite-layer nickelates. Further investigations are necessary to clarify the role of the rare-earth 5$d$ Fermi surface (FS) pocket and strain in the emergence of charge order.  

In summary, our RIXS data showcase the orbital configuration and magnetic excitations of early rare-earth infinite-layer nickelate thin films with a SrTiO$_3$ capping layer. The observed universal orbital configuration and magnetic features indicate a common behavior among these nickelates. Interestingly, we only observed a signature of charge order in LaNiO$_2$, but not in PrNiO$_2$ and NdNiO$_2$, calling for further investigations. As an outlook, it would be interesting to investigate infinite-layer nickelates with high-Z rare-earth elements, which can further examine the universality of the electronic and magnetic structures, as well as the emergence of other instabilities, as predicted by some theories \cite{Zhang2023,Subedi2023}. 

\begin{acknowledgments}
This work is supported by the U.S. Department of Energy (DOE), Office of Science, Basic Energy Sciences, Materials Sciences and Engineering Division, under contract DE-AC02-76SF00515. We acknowledge the Gordon and Betty Moore Foundation’s Emergent Phenomena in Quantum Systems Initiative through grant number GBMF9072 for synthesis equipment. We acknowledge Diamond Light Source for time on beamline I21-RIXS under Proposal NT25165 and MM25598. This research also used resources of the Advanced Light Source, a U.S. DOE Office of Science User Facility under contract no. DE-AC02-05CH11231. B.H.G. acknowledges support by the Department of Defense Air Force Office of Scientific Research (No. FA 9550-16-1-0305). This work made use of the Cornell Center for Materials Research (CCMR) Shared Facilities, which are supported through the NSF MRSEC Program (No. DMR-1719875). The FEI Titan Themis 300 was acquired through No. NSF-MRI-1429155, with additional support from Cornell University, the Weill Institute, and the Kavli Institute at Cornell. The Thermo Fisher Helios G4 UX FIB was acquired with support by NSF No. DMR-1539918. The Thermo Fisher Spectra 300 X-CFEG was acquired with support from PARADIM, an NSF MIP (DMR-2039380), and Cornell University.
\end{acknowledgments}


\begin{thebibliography}{55}%
\makeatletter
\providecommand \@ifxundefined [1]{%
 \@ifx{#1\undefined}
}%
\providecommand \@ifnum [1]{%
 \ifnum #1\expandafter \@firstoftwo
 \else \expandafter \@secondoftwo
 \fi
}%
\providecommand \@ifx [1]{%
 \ifx #1\expandafter \@firstoftwo
 \else \expandafter \@secondoftwo
 \fi
}%
\providecommand \natexlab [1]{#1}%
\providecommand \enquote  [1]{``#1''}%
\providecommand \bibnamefont  [1]{#1}%
\providecommand \bibfnamefont [1]{#1}%
\providecommand \citenamefont [1]{#1}%
\providecommand \href@noop [0]{\@secondoftwo}%
\providecommand \href [0]{\begingroup \@sanitize@url \@href}%
\providecommand \@href[1]{\@@startlink{#1}\@@href}%
\providecommand \@@href[1]{\endgroup#1\@@endlink}%
\providecommand \@sanitize@url [0]{\catcode `\\12\catcode `\$12\catcode `\&12\catcode `\#12\catcode `\^12\catcode `\_12\catcode `\%12\relax}%
\providecommand \@@startlink[1]{}%
\providecommand \@@endlink[0]{}%
\providecommand \url  [0]{\begingroup\@sanitize@url \@url }%
\providecommand \@url [1]{\endgroup\@href {#1}{\urlprefix }}%
\providecommand \urlprefix  [0]{URL }%
\providecommand \Eprint [0]{\href }%
\providecommand \doibase [0]{https://doi.org/}%
\providecommand \selectlanguage [0]{\@gobble}%
\providecommand \bibinfo  [0]{\@secondoftwo}%
\providecommand \bibfield  [0]{\@secondoftwo}%
\providecommand \translation [1]{[#1]}%
\providecommand \BibitemOpen [0]{}%
\providecommand \bibitemStop [0]{}%
\providecommand \bibitemNoStop [0]{.\EOS\space}%
\providecommand \EOS [0]{\spacefactor3000\relax}%
\providecommand \BibitemShut  [1]{\csname bibitem#1\endcsname}%
\let\auto@bib@innerbib\@empty
\bibitem [{\citenamefont {Li}\ \emph {et~al.}(2019)\citenamefont {Li}, \citenamefont {Lee}, \citenamefont {Wang}, \citenamefont {Osada}, \citenamefont {Crossley}, \citenamefont {Lee}, \citenamefont {Cui}, \citenamefont {Hikita},\ and\ \citenamefont {Hwang}}]{Li2019}%
  \BibitemOpen
  \bibfield  {author} {\bibinfo {author} {\bibfnamefont {D.}~\bibnamefont {Li}}, \bibinfo {author} {\bibfnamefont {K.}~\bibnamefont {Lee}}, \bibinfo {author} {\bibfnamefont {B.~Y.}\ \bibnamefont {Wang}}, \bibinfo {author} {\bibfnamefont {M.}~\bibnamefont {Osada}}, \bibinfo {author} {\bibfnamefont {S.}~\bibnamefont {Crossley}}, \bibinfo {author} {\bibfnamefont {H.~R.}\ \bibnamefont {Lee}}, \bibinfo {author} {\bibfnamefont {Y.}~\bibnamefont {Cui}}, \bibinfo {author} {\bibfnamefont {Y.}~\bibnamefont {Hikita}},\ and\ \bibinfo {author} {\bibfnamefont {H.~Y.}\ \bibnamefont {Hwang}},\ }\bibfield  {title} {\bibinfo {title} {Superconductivity in an infinite-layer nickelate},\ }\href {https://doi.org/10.1038/s41586-019-1496-5} {\bibfield  {journal} {\bibinfo  {journal} {Nature}\ }\textbf {\bibinfo {volume} {572}},\ \bibinfo {pages} {624} (\bibinfo {year} {2019})}\BibitemShut {NoStop}%
\bibitem [{\citenamefont {Anisimov}\ \emph {et~al.}(1999)\citenamefont {Anisimov}, \citenamefont {Bukhvalov},\ and\ \citenamefont {Rice}}]{Anisimov1999}%
  \BibitemOpen
  \bibfield  {author} {\bibinfo {author} {\bibfnamefont {V.~I.}\ \bibnamefont {Anisimov}}, \bibinfo {author} {\bibfnamefont {D.}~\bibnamefont {Bukhvalov}},\ and\ \bibinfo {author} {\bibfnamefont {T.~M.}\ \bibnamefont {Rice}},\ }\bibfield  {title} {\bibinfo {title} {{Electronic structure of possible nickelate analogs to the cuprates}},\ }\href {https://doi.org/10.1103/PhysRevB.59.7901} {\bibfield  {journal} {\bibinfo  {journal} {Phys. Rev. B}\ }\textbf {\bibinfo {volume} {59}},\ \bibinfo {pages} {7901} (\bibinfo {year} {1999})}\BibitemShut {NoStop}%
\bibitem [{\citenamefont {Lee}\ and\ \citenamefont {Pickett}(2004)}]{Lee2004}%
  \BibitemOpen
  \bibfield  {author} {\bibinfo {author} {\bibfnamefont {K.-W.}\ \bibnamefont {Lee}}\ and\ \bibinfo {author} {\bibfnamefont {W.~E.}\ \bibnamefont {Pickett}},\ }\bibfield  {title} {\bibinfo {title} {{Infinite-layer $\mathrm{La}\mathrm{Ni}{\mathrm{O}}_{2}$: ${\mathrm{Ni}}^{1+}$ is not ${\mathrm{Cu}}^{2+}$}},\ }\href {https://doi.org/10.1103/PhysRevB.70.165109} {\bibfield  {journal} {\bibinfo  {journal} {Phys. Rev. B}\ }\textbf {\bibinfo {volume} {70}},\ \bibinfo {pages} {165109} (\bibinfo {year} {2004})}\BibitemShut {NoStop}%
\bibitem [{\citenamefont {Li}\ \emph {et~al.}(2020)\citenamefont {Li}, \citenamefont {Wang}, \citenamefont {Lee}, \citenamefont {Harvey}, \citenamefont {Osada}, \citenamefont {Goodge}, \citenamefont {Kourkoutis},\ and\ \citenamefont {Hwang}}]{Li2020}%
  \BibitemOpen
  \bibfield  {author} {\bibinfo {author} {\bibfnamefont {D.}~\bibnamefont {Li}}, \bibinfo {author} {\bibfnamefont {B.~Y.}\ \bibnamefont {Wang}}, \bibinfo {author} {\bibfnamefont {K.}~\bibnamefont {Lee}}, \bibinfo {author} {\bibfnamefont {S.~P.}\ \bibnamefont {Harvey}}, \bibinfo {author} {\bibfnamefont {M.}~\bibnamefont {Osada}}, \bibinfo {author} {\bibfnamefont {B.~H.}\ \bibnamefont {Goodge}}, \bibinfo {author} {\bibfnamefont {L.~F.}\ \bibnamefont {Kourkoutis}},\ and\ \bibinfo {author} {\bibfnamefont {H.~Y.}\ \bibnamefont {Hwang}},\ }\bibfield  {title} {\bibinfo {title} {{Superconducting dome in ${\mathrm{Nd}}_{1\ensuremath{-}x}{\mathrm{Sr}}_{x}{\mathrm{NiO}}_{2}$ infinite layer films}},\ }\href {https://doi.org/10.1103/PhysRevLett.125.027001} {\bibfield  {journal} {\bibinfo  {journal} {Phys. Rev. Lett.}\ }\textbf {\bibinfo {volume} {125}},\ \bibinfo {pages} {027001} (\bibinfo {year} {2020})}\BibitemShut {NoStop}%
\bibitem [{\citenamefont {Zeng}\ \emph {et~al.}(2020)\citenamefont {Zeng}, \citenamefont {Tang}, \citenamefont {Yin}, \citenamefont {Li}, \citenamefont {Li}, \citenamefont {Huang}, \citenamefont {Hu}, \citenamefont {Liu}, \citenamefont {Omar}, \citenamefont {Jani}, \citenamefont {Lim}, \citenamefont {Han}, \citenamefont {Wan}, \citenamefont {Yang}, \citenamefont {Pennycook}, \citenamefont {Wee},\ and\ \citenamefont {Ariando}}]{Zeng2020}%
  \BibitemOpen
  \bibfield  {author} {\bibinfo {author} {\bibfnamefont {S.}~\bibnamefont {Zeng}}, \bibinfo {author} {\bibfnamefont {C.~S.}\ \bibnamefont {Tang}}, \bibinfo {author} {\bibfnamefont {X.}~\bibnamefont {Yin}}, \bibinfo {author} {\bibfnamefont {C.}~\bibnamefont {Li}}, \bibinfo {author} {\bibfnamefont {M.}~\bibnamefont {Li}}, \bibinfo {author} {\bibfnamefont {Z.}~\bibnamefont {Huang}}, \bibinfo {author} {\bibfnamefont {J.}~\bibnamefont {Hu}}, \bibinfo {author} {\bibfnamefont {W.}~\bibnamefont {Liu}}, \bibinfo {author} {\bibfnamefont {G.~J.}\ \bibnamefont {Omar}}, \bibinfo {author} {\bibfnamefont {H.}~\bibnamefont {Jani}}, \bibinfo {author} {\bibfnamefont {Z.~S.}\ \bibnamefont {Lim}}, \bibinfo {author} {\bibfnamefont {K.}~\bibnamefont {Han}}, \bibinfo {author} {\bibfnamefont {D.}~\bibnamefont {Wan}}, \bibinfo {author} {\bibfnamefont {P.}~\bibnamefont {Yang}}, \bibinfo {author} {\bibfnamefont {S.~J.}\ \bibnamefont {Pennycook}}, \bibinfo {author} {\bibfnamefont {A.~T.~S.}\ \bibnamefont {Wee}},\ and\ \bibinfo {author}
  {\bibfnamefont {A.}~\bibnamefont {Ariando}},\ }\bibfield  {title} {\bibinfo {title} {{Phase diagram and superconducting dome of infinite-layer ${\mathrm{Nd}}_{1\ensuremath{-}x}{\mathrm{Sr}}_{x}{\mathrm{NiO}}_{2}$ thin films}},\ }\href {https://doi.org/10.1103/PhysRevLett.125.147003} {\bibfield  {journal} {\bibinfo  {journal} {Phys. Rev. Lett.}\ }\textbf {\bibinfo {volume} {125}},\ \bibinfo {pages} {147003} (\bibinfo {year} {2020})}\BibitemShut {NoStop}%
\bibitem [{\citenamefont {Lee}\ \emph {et~al.}(2023)\citenamefont {Lee}, \citenamefont {Wang}, \citenamefont {Osada}, \citenamefont {Goodge}, \citenamefont {Wang}, \citenamefont {Lee}, \citenamefont {Harvey}, \citenamefont {Kim}, \citenamefont {Yu}, \citenamefont {Murthy}, \citenamefont {Raghu}, \citenamefont {Kourkoutis},\ and\ \citenamefont {Hwang}}]{Lee2023}%
  \BibitemOpen
  \bibfield  {author} {\bibinfo {author} {\bibfnamefont {K.}~\bibnamefont {Lee}}, \bibinfo {author} {\bibfnamefont {B.~Y.}\ \bibnamefont {Wang}}, \bibinfo {author} {\bibfnamefont {M.}~\bibnamefont {Osada}}, \bibinfo {author} {\bibfnamefont {B.~H.}\ \bibnamefont {Goodge}}, \bibinfo {author} {\bibfnamefont {T.~C.}\ \bibnamefont {Wang}}, \bibinfo {author} {\bibfnamefont {Y.}~\bibnamefont {Lee}}, \bibinfo {author} {\bibfnamefont {S.}~\bibnamefont {Harvey}}, \bibinfo {author} {\bibfnamefont {W.~J.}\ \bibnamefont {Kim}}, \bibinfo {author} {\bibfnamefont {Y.}~\bibnamefont {Yu}}, \bibinfo {author} {\bibfnamefont {C.}~\bibnamefont {Murthy}}, \bibinfo {author} {\bibfnamefont {S.}~\bibnamefont {Raghu}}, \bibinfo {author} {\bibfnamefont {L.~F.}\ \bibnamefont {Kourkoutis}},\ and\ \bibinfo {author} {\bibfnamefont {H.~Y.}\ \bibnamefont {Hwang}},\ }\bibfield  {title} {\bibinfo {title} {Linear-in-temperature resistivity for optimally superconducting (nd,sr)nio2},\ }\href {https://doi.org/10.1038/s41586-023-06129-x} {\bibfield
   {journal} {\bibinfo  {journal} {Nature}\ }\textbf {\bibinfo {volume} {619}},\ \bibinfo {pages} {288} (\bibinfo {year} {2023})}\BibitemShut {NoStop}%
\bibitem [{\citenamefont {Hepting}\ \emph {et~al.}(2018)\citenamefont {Hepting}, \citenamefont {Chaix}, \citenamefont {Huang}, \citenamefont {Fumagalli}, \citenamefont {Peng}, \citenamefont {Moritz}, \citenamefont {Kummer}, \citenamefont {Brookes}, \citenamefont {Lee}, \citenamefont {Hashimoto}, \citenamefont {Sarkar}, \citenamefont {He}, \citenamefont {Rotundu}, \citenamefont {Lee}, \citenamefont {Greene}, \citenamefont {Braicovich}, \citenamefont {Ghiringhelli}, \citenamefont {Shen}, \citenamefont {Devereaux},\ and\ \citenamefont {Lee}}]{Hepting2018}%
  \BibitemOpen
  \bibfield  {author} {\bibinfo {author} {\bibfnamefont {M.}~\bibnamefont {Hepting}}, \bibinfo {author} {\bibfnamefont {L.}~\bibnamefont {Chaix}}, \bibinfo {author} {\bibfnamefont {E.~W.}\ \bibnamefont {Huang}}, \bibinfo {author} {\bibfnamefont {R.}~\bibnamefont {Fumagalli}}, \bibinfo {author} {\bibfnamefont {Y.~Y.}\ \bibnamefont {Peng}}, \bibinfo {author} {\bibfnamefont {B.}~\bibnamefont {Moritz}}, \bibinfo {author} {\bibfnamefont {K.}~\bibnamefont {Kummer}}, \bibinfo {author} {\bibfnamefont {N.~B.}\ \bibnamefont {Brookes}}, \bibinfo {author} {\bibfnamefont {W.~C.}\ \bibnamefont {Lee}}, \bibinfo {author} {\bibfnamefont {M.}~\bibnamefont {Hashimoto}}, \bibinfo {author} {\bibfnamefont {T.}~\bibnamefont {Sarkar}}, \bibinfo {author} {\bibfnamefont {J.-F.}\ \bibnamefont {He}}, \bibinfo {author} {\bibfnamefont {C.~R.}\ \bibnamefont {Rotundu}}, \bibinfo {author} {\bibfnamefont {Y.~S.}\ \bibnamefont {Lee}}, \bibinfo {author} {\bibfnamefont {R.~L.}\ \bibnamefont {Greene}}, \bibinfo {author} {\bibfnamefont
  {L.}~\bibnamefont {Braicovich}}, \bibinfo {author} {\bibfnamefont {G.}~\bibnamefont {Ghiringhelli}}, \bibinfo {author} {\bibfnamefont {Z.~X.}\ \bibnamefont {Shen}}, \bibinfo {author} {\bibfnamefont {T.~P.}\ \bibnamefont {Devereaux}},\ and\ \bibinfo {author} {\bibfnamefont {W.~S.}\ \bibnamefont {Lee}},\ }\bibfield  {title} {\bibinfo {title} {{Three-dimensional collective charge excitations in electron-doped copper oxide superconductors}},\ }\href {https://doi.org/10.1038/s41586-018-0648-3} {\bibfield  {journal} {\bibinfo  {journal} {Nature}\ }\textbf {\bibinfo {volume} {563}},\ \bibinfo {pages} {374} (\bibinfo {year} {2018})}\BibitemShut {NoStop}%
\bibitem [{\citenamefont {Rossi}\ \emph {et~al.}(2021)\citenamefont {Rossi}, \citenamefont {Lu}, \citenamefont {Nag}, \citenamefont {Li}, \citenamefont {Osada}, \citenamefont {Lee}, \citenamefont {Wang}, \citenamefont {Agrestini}, \citenamefont {Garcia-Fernandez}, \citenamefont {Kas}, \citenamefont {Chuang}, \citenamefont {Shen}, \citenamefont {Hwang}, \citenamefont {Moritz}, \citenamefont {Zhou}, \citenamefont {Devereaux},\ and\ \citenamefont {Lee}}]{Rossi2021}%
  \BibitemOpen
  \bibfield  {author} {\bibinfo {author} {\bibfnamefont {M.}~\bibnamefont {Rossi}}, \bibinfo {author} {\bibfnamefont {H.}~\bibnamefont {Lu}}, \bibinfo {author} {\bibfnamefont {A.}~\bibnamefont {Nag}}, \bibinfo {author} {\bibfnamefont {D.}~\bibnamefont {Li}}, \bibinfo {author} {\bibfnamefont {M.}~\bibnamefont {Osada}}, \bibinfo {author} {\bibfnamefont {K.}~\bibnamefont {Lee}}, \bibinfo {author} {\bibfnamefont {B.~Y.}\ \bibnamefont {Wang}}, \bibinfo {author} {\bibfnamefont {S.}~\bibnamefont {Agrestini}}, \bibinfo {author} {\bibfnamefont {M.}~\bibnamefont {Garcia-Fernandez}}, \bibinfo {author} {\bibfnamefont {J.~J.}\ \bibnamefont {Kas}}, \bibinfo {author} {\bibfnamefont {Y.-D.}\ \bibnamefont {Chuang}}, \bibinfo {author} {\bibfnamefont {Z.~X.}\ \bibnamefont {Shen}}, \bibinfo {author} {\bibfnamefont {H.~Y.}\ \bibnamefont {Hwang}}, \bibinfo {author} {\bibfnamefont {B.}~\bibnamefont {Moritz}}, \bibinfo {author} {\bibfnamefont {K.-J.}\ \bibnamefont {Zhou}}, \bibinfo {author} {\bibfnamefont {T.~P.}\ \bibnamefont
  {Devereaux}},\ and\ \bibinfo {author} {\bibfnamefont {W.~S.}\ \bibnamefont {Lee}},\ }\bibfield  {title} {\bibinfo {title} {Orbital and spin character of doped carriers in infinite-layer nickelates},\ }\href {https://doi.org/10.1103/PhysRevB.104.L220505} {\bibfield  {journal} {\bibinfo  {journal} {Phys. Rev. B}\ }\textbf {\bibinfo {volume} {104}},\ \bibinfo {pages} {L220505} (\bibinfo {year} {2021})}\BibitemShut {NoStop}%
\bibitem [{\citenamefont {Goodge}\ \emph {et~al.}(2021)\citenamefont {Goodge}, \citenamefont {Li}, \citenamefont {Lee}, \citenamefont {Osada}, \citenamefont {Wang}, \citenamefont {Sawatzky}, \citenamefont {Hwang},\ and\ \citenamefont {Kourkoutis}}]{Goodge2021}%
  \BibitemOpen
  \bibfield  {author} {\bibinfo {author} {\bibfnamefont {B.~H.}\ \bibnamefont {Goodge}}, \bibinfo {author} {\bibfnamefont {D.}~\bibnamefont {Li}}, \bibinfo {author} {\bibfnamefont {K.}~\bibnamefont {Lee}}, \bibinfo {author} {\bibfnamefont {M.}~\bibnamefont {Osada}}, \bibinfo {author} {\bibfnamefont {B.~Y.}\ \bibnamefont {Wang}}, \bibinfo {author} {\bibfnamefont {G.~A.}\ \bibnamefont {Sawatzky}}, \bibinfo {author} {\bibfnamefont {H.~Y.}\ \bibnamefont {Hwang}},\ and\ \bibinfo {author} {\bibfnamefont {L.~F.}\ \bibnamefont {Kourkoutis}},\ }\bibfield  {title} {\bibinfo {title} {Doping evolution of the mott{\textendash}hubbard landscape in infinite-layer nickelates},\ }\bibfield  {journal} {\bibinfo  {journal} {Proc. Natl. Acad. Sci. USA}\ }\textbf {\bibinfo {volume} {118}},\ \href {https://doi.org/10.1073/pnas.2007683118} {10.1073/pnas.2007683118} (\bibinfo {year} {2021})\BibitemShut {NoStop}%
\bibitem [{\citenamefont {Rossi}\ \emph {et~al.}(2022)\citenamefont {Rossi}, \citenamefont {Osada}, \citenamefont {Choi}, \citenamefont {Agrestini}, \citenamefont {Jost}, \citenamefont {Lee}, \citenamefont {Lu}, \citenamefont {Wang}, \citenamefont {Lee}, \citenamefont {Nag}, \citenamefont {Chuang}, \citenamefont {Kuo}, \citenamefont {Lee}, \citenamefont {Moritz}, \citenamefont {Devereaux}, \citenamefont {Shen}, \citenamefont {Lee}, \citenamefont {Zhou}, \citenamefont {Hwang},\ and\ \citenamefont {Lee}}]{Rossi2022}%
  \BibitemOpen
  \bibfield  {author} {\bibinfo {author} {\bibfnamefont {M.}~\bibnamefont {Rossi}}, \bibinfo {author} {\bibfnamefont {M.}~\bibnamefont {Osada}}, \bibinfo {author} {\bibfnamefont {J.}~\bibnamefont {Choi}}, \bibinfo {author} {\bibfnamefont {S.}~\bibnamefont {Agrestini}}, \bibinfo {author} {\bibfnamefont {D.}~\bibnamefont {Jost}}, \bibinfo {author} {\bibfnamefont {Y.}~\bibnamefont {Lee}}, \bibinfo {author} {\bibfnamefont {H.}~\bibnamefont {Lu}}, \bibinfo {author} {\bibfnamefont {B.~Y.}\ \bibnamefont {Wang}}, \bibinfo {author} {\bibfnamefont {K.}~\bibnamefont {Lee}}, \bibinfo {author} {\bibfnamefont {A.}~\bibnamefont {Nag}}, \bibinfo {author} {\bibfnamefont {Y.-D.}\ \bibnamefont {Chuang}}, \bibinfo {author} {\bibfnamefont {C.-T.}\ \bibnamefont {Kuo}}, \bibinfo {author} {\bibfnamefont {S.-J.}\ \bibnamefont {Lee}}, \bibinfo {author} {\bibfnamefont {B.}~\bibnamefont {Moritz}}, \bibinfo {author} {\bibfnamefont {T.~P.}\ \bibnamefont {Devereaux}}, \bibinfo {author} {\bibfnamefont {Z.-X.}\ \bibnamefont {Shen}}, \bibinfo
  {author} {\bibfnamefont {J.-S.}\ \bibnamefont {Lee}}, \bibinfo {author} {\bibfnamefont {K.-J.}\ \bibnamefont {Zhou}}, \bibinfo {author} {\bibfnamefont {H.~Y.}\ \bibnamefont {Hwang}},\ and\ \bibinfo {author} {\bibfnamefont {W.-S.}\ \bibnamefont {Lee}},\ }\bibfield  {title} {\bibinfo {title} {A broken translational symmetry state in an infinite-layer nickelate},\ }\href {https://doi.org/10.1038/s41567-022-01660-6} {\bibfield  {journal} {\bibinfo  {journal} {Nature Physics}\ }\textbf {\bibinfo {volume} {18}},\ \bibinfo {pages} {869} (\bibinfo {year} {2022})}\BibitemShut {NoStop}%
\bibitem [{\citenamefont {Osada}\ \emph {et~al.}(2020{\natexlab{a}})\citenamefont {Osada}, \citenamefont {Wang}, \citenamefont {Lee}, \citenamefont {Li},\ and\ \citenamefont {Hwang}}]{Osada2020}%
  \BibitemOpen
  \bibfield  {author} {\bibinfo {author} {\bibfnamefont {M.}~\bibnamefont {Osada}}, \bibinfo {author} {\bibfnamefont {B.~Y.}\ \bibnamefont {Wang}}, \bibinfo {author} {\bibfnamefont {K.}~\bibnamefont {Lee}}, \bibinfo {author} {\bibfnamefont {D.}~\bibnamefont {Li}},\ and\ \bibinfo {author} {\bibfnamefont {H.~Y.}\ \bibnamefont {Hwang}},\ }\bibfield  {title} {\bibinfo {title} {Phase diagram of infinite layer praseodymium nickelate {Pr$_{1-x}$Sr$_x$NiO$_2$} thin films},\ }\href {https://doi.org/10.1103/PhysRevMaterials.4.121801} {\bibfield  {journal} {\bibinfo  {journal} {Phys. Rev. Mater.}\ }\textbf {\bibinfo {volume} {4}},\ \bibinfo {pages} {121801} (\bibinfo {year} {2020}{\natexlab{a}})}\BibitemShut {NoStop}%
\bibitem [{\citenamefont {Osada}\ \emph {et~al.}(2021)\citenamefont {Osada}, \citenamefont {Wang}, \citenamefont {Goodge}, \citenamefont {Harvey}, \citenamefont {Lee}, \citenamefont {Li}, \citenamefont {Kourkoutis},\ and\ \citenamefont {Hwang}}]{Osada2021}%
  \BibitemOpen
  \bibfield  {author} {\bibinfo {author} {\bibfnamefont {M.}~\bibnamefont {Osada}}, \bibinfo {author} {\bibfnamefont {B.~Y.}\ \bibnamefont {Wang}}, \bibinfo {author} {\bibfnamefont {B.~H.}\ \bibnamefont {Goodge}}, \bibinfo {author} {\bibfnamefont {S.~P.}\ \bibnamefont {Harvey}}, \bibinfo {author} {\bibfnamefont {K.}~\bibnamefont {Lee}}, \bibinfo {author} {\bibfnamefont {D.}~\bibnamefont {Li}}, \bibinfo {author} {\bibfnamefont {L.~F.}\ \bibnamefont {Kourkoutis}},\ and\ \bibinfo {author} {\bibfnamefont {H.~Y.}\ \bibnamefont {Hwang}},\ }\bibfield  {title} {\bibinfo {title} {{Nickelate superconductivity without sare-earth magnetism: (La,Sr)NiO$_2$}},\ }\href {https://doi.org/https://doi.org/10.1002/adma.202104083} {\bibfield  {journal} {\bibinfo  {journal} {Advanced Materials}\ }\textbf {\bibinfo {volume} {33}},\ \bibinfo {pages} {2104083} (\bibinfo {year} {2021})},\ \Eprint {https://arxiv.org/abs/https://onlinelibrary.wiley.com/doi/pdf/10.1002/adma.202104083}
  {https://onlinelibrary.wiley.com/doi/pdf/10.1002/adma.202104083} \BibitemShut {NoStop}%
\bibitem [{\citenamefont {Harvey}\ \emph {et~al.}(2022)\citenamefont {Harvey}, \citenamefont {Wang}, \citenamefont {Fowlie}, \citenamefont {Osada}, \citenamefont {Lee}, \citenamefont {Lee}, \citenamefont {Li},\ and\ \citenamefont {Hwang}}]{harvey2022}%
  \BibitemOpen
  \bibfield  {author} {\bibinfo {author} {\bibfnamefont {S.~P.}\ \bibnamefont {Harvey}}, \bibinfo {author} {\bibfnamefont {B.~Y.}\ \bibnamefont {Wang}}, \bibinfo {author} {\bibfnamefont {J.}~\bibnamefont {Fowlie}}, \bibinfo {author} {\bibfnamefont {M.}~\bibnamefont {Osada}}, \bibinfo {author} {\bibfnamefont {K.}~\bibnamefont {Lee}}, \bibinfo {author} {\bibfnamefont {Y.}~\bibnamefont {Lee}}, \bibinfo {author} {\bibfnamefont {D.}~\bibnamefont {Li}},\ and\ \bibinfo {author} {\bibfnamefont {H.~Y.}\ \bibnamefont {Hwang}},\ }\href@noop {} {\bibinfo {title} {Evidence for nodal superconductivity in infinite-layer nickelates}} (\bibinfo {year} {2022}),\ \Eprint {https://arxiv.org/abs/2201.12971} {arXiv:2201.12971 [cond-mat.supr-con]} \BibitemShut {NoStop}%
\bibitem [{\citenamefont {Wang}\ \emph {et~al.}(2023)\citenamefont {Wang}, \citenamefont {Wang}, \citenamefont {Hsu}, \citenamefont {Osada}, \citenamefont {Lee}, \citenamefont {Jia}, \citenamefont {Duffy}, \citenamefont {Li}, \citenamefont {Fowlie}, \citenamefont {Beasley}, \citenamefont {Devereaux}, \citenamefont {Fisher}, \citenamefont {Hussey},\ and\ \citenamefont {Hwang}}]{Wang2023}%
  \BibitemOpen
  \bibfield  {author} {\bibinfo {author} {\bibfnamefont {B.~Y.}\ \bibnamefont {Wang}}, \bibinfo {author} {\bibfnamefont {T.~C.}\ \bibnamefont {Wang}}, \bibinfo {author} {\bibfnamefont {Y.-T.}\ \bibnamefont {Hsu}}, \bibinfo {author} {\bibfnamefont {M.}~\bibnamefont {Osada}}, \bibinfo {author} {\bibfnamefont {K.}~\bibnamefont {Lee}}, \bibinfo {author} {\bibfnamefont {C.}~\bibnamefont {Jia}}, \bibinfo {author} {\bibfnamefont {C.}~\bibnamefont {Duffy}}, \bibinfo {author} {\bibfnamefont {D.}~\bibnamefont {Li}}, \bibinfo {author} {\bibfnamefont {J.}~\bibnamefont {Fowlie}}, \bibinfo {author} {\bibfnamefont {M.~R.}\ \bibnamefont {Beasley}}, \bibinfo {author} {\bibfnamefont {T.~P.}\ \bibnamefont {Devereaux}}, \bibinfo {author} {\bibfnamefont {I.~R.}\ \bibnamefont {Fisher}}, \bibinfo {author} {\bibfnamefont {N.~E.}\ \bibnamefont {Hussey}},\ and\ \bibinfo {author} {\bibfnamefont {H.~Y.}\ \bibnamefont {Hwang}},\ }\bibfield  {title} {\bibinfo {title} {Effects of rare-earth magnetism on the superconducting upper critical
  field in infinite-layer nickelates},\ }\href {https://doi.org/10.1126/sciadv.adf6655} {\bibfield  {journal} {\bibinfo  {journal} {Science Advances}\ }\textbf {\bibinfo {volume} {9}},\ \bibinfo {pages} {eadf6655} (\bibinfo {year} {2023})},\ \Eprint {https://arxiv.org/abs/https://www.science.org/doi/pdf/10.1126/sciadv.adf6655} {https://www.science.org/doi/pdf/10.1126/sciadv.adf6655} \BibitemShut {NoStop}%
\bibitem [{\citenamefont {Nomura}\ \emph {et~al.}(2019)\citenamefont {Nomura}, \citenamefont {Hirayama}, \citenamefont {Tadano}, \citenamefont {Yoshimoto}, \citenamefont {Nakamura},\ and\ \citenamefont {Arita}}]{Nomura2019}%
  \BibitemOpen
  \bibfield  {author} {\bibinfo {author} {\bibfnamefont {Y.}~\bibnamefont {Nomura}}, \bibinfo {author} {\bibfnamefont {M.}~\bibnamefont {Hirayama}}, \bibinfo {author} {\bibfnamefont {T.}~\bibnamefont {Tadano}}, \bibinfo {author} {\bibfnamefont {Y.}~\bibnamefont {Yoshimoto}}, \bibinfo {author} {\bibfnamefont {K.}~\bibnamefont {Nakamura}},\ and\ \bibinfo {author} {\bibfnamefont {R.}~\bibnamefont {Arita}},\ }\bibfield  {title} {\bibinfo {title} {{Formation of a two-dimensional single-component correlated electron system and band engineering in the nickelate superconductor ${\mathrm{NdNiO}}_{2}$}},\ }\href {https://doi.org/10.1103/PhysRevB.100.205138} {\bibfield  {journal} {\bibinfo  {journal} {Phys. Rev. B}\ }\textbf {\bibinfo {volume} {100}},\ \bibinfo {pages} {205138} (\bibinfo {year} {2019})}\BibitemShut {NoStop}%
\bibitem [{\citenamefont {Adhikary}\ \emph {et~al.}(2020)\citenamefont {Adhikary}, \citenamefont {Bandyopadhyay}, \citenamefont {Das}, \citenamefont {Dasgupta},\ and\ \citenamefont {Saha-Dasgupta}}]{Adhikary2020}%
  \BibitemOpen
  \bibfield  {author} {\bibinfo {author} {\bibfnamefont {P.}~\bibnamefont {Adhikary}}, \bibinfo {author} {\bibfnamefont {S.}~\bibnamefont {Bandyopadhyay}}, \bibinfo {author} {\bibfnamefont {T.}~\bibnamefont {Das}}, \bibinfo {author} {\bibfnamefont {I.}~\bibnamefont {Dasgupta}},\ and\ \bibinfo {author} {\bibfnamefont {T.}~\bibnamefont {Saha-Dasgupta}},\ }\bibfield  {title} {\bibinfo {title} {{Orbital-selective superconductivity in a two-band model of infinite-layer nickelates}},\ }\href {https://doi.org/10.1103/PhysRevB.102.100501} {\bibfield  {journal} {\bibinfo  {journal} {Phys. Rev. B}\ }\textbf {\bibinfo {volume} {102}},\ \bibinfo {pages} {100501(R)} (\bibinfo {year} {2020})}\BibitemShut {NoStop}%
\bibitem [{\citenamefont {Botana}\ and\ \citenamefont {Norman}(2020)}]{Botana2020}%
  \BibitemOpen
  \bibfield  {author} {\bibinfo {author} {\bibfnamefont {A.~S.}\ \bibnamefont {Botana}}\ and\ \bibinfo {author} {\bibfnamefont {M.~R.}\ \bibnamefont {Norman}},\ }\bibfield  {title} {\bibinfo {title} {{Similarities and Differences between ${\mathrm{LaNiO}}_{2}$ and ${\mathrm{CaCuO}}_{2}$ and Implications for Superconductivity}},\ }\href {https://doi.org/10.1103/PhysRevX.10.011024} {\bibfield  {journal} {\bibinfo  {journal} {Phys. Rev. X}\ }\textbf {\bibinfo {volume} {10}},\ \bibinfo {pages} {011024} (\bibinfo {year} {2020})}\BibitemShut {NoStop}%
\bibitem [{\citenamefont {Gu}\ \emph {et~al.}(2020)\citenamefont {Gu}, \citenamefont {Zhu}, \citenamefont {Wang}, \citenamefont {Hu},\ and\ \citenamefont {Chen}}]{Gu2020a}%
  \BibitemOpen
  \bibfield  {author} {\bibinfo {author} {\bibfnamefont {Y.}~\bibnamefont {Gu}}, \bibinfo {author} {\bibfnamefont {S.}~\bibnamefont {Zhu}}, \bibinfo {author} {\bibfnamefont {X.}~\bibnamefont {Wang}}, \bibinfo {author} {\bibfnamefont {J.}~\bibnamefont {Hu}},\ and\ \bibinfo {author} {\bibfnamefont {H.}~\bibnamefont {Chen}},\ }\bibfield  {title} {\bibinfo {title} {{A substantial hybridization between correlated Ni-d orbital and itinerant electrons in infinite-layer nickelates}},\ }\href {https://doi.org/10.1038/s42005-020-0347-x} {\bibfield  {journal} {\bibinfo  {journal} {Commun. Phys.}\ }\textbf {\bibinfo {volume} {3}},\ \bibinfo {pages} {84} (\bibinfo {year} {2020})}\BibitemShut {NoStop}%
\bibitem [{\citenamefont {Kapeghian}\ and\ \citenamefont {Botana}(2020)}]{Kapeghian2020}%
  \BibitemOpen
  \bibfield  {author} {\bibinfo {author} {\bibfnamefont {J.}~\bibnamefont {Kapeghian}}\ and\ \bibinfo {author} {\bibfnamefont {A.~S.}\ \bibnamefont {Botana}},\ }\bibfield  {title} {\bibinfo {title} {{Electronic structure and magnetism in infinite-layer nickelates $R{\mathrm{NiO}}_{2}$ ($R=\mathrm{La}\text{\ensuremath{-}}\mathrm{Lu}$)}},\ }\href {https://doi.org/10.1103/PhysRevB.102.205130} {\bibfield  {journal} {\bibinfo  {journal} {Phys. Rev. B}\ }\textbf {\bibinfo {volume} {102}},\ \bibinfo {pages} {205130} (\bibinfo {year} {2020})}\BibitemShut {NoStop}%
\bibitem [{\citenamefont {Lechermann}(2020)}]{Lechermann2020}%
  \BibitemOpen
  \bibfield  {author} {\bibinfo {author} {\bibfnamefont {F.}~\bibnamefont {Lechermann}},\ }\bibfield  {title} {\bibinfo {title} {{Late transition metal oxides with infinite-layer structure: Nickelates versus cuprates}},\ }\href {https://doi.org/10.1103/PhysRevB.101.081110} {\bibfield  {journal} {\bibinfo  {journal} {Phys. Rev. B}\ }\textbf {\bibinfo {volume} {101}},\ \bibinfo {pages} {081110} (\bibinfo {year} {2020})}\BibitemShut {NoStop}%
\bibitem [{\citenamefont {Leonov}\ \emph {et~al.}(2020)\citenamefont {Leonov}, \citenamefont {Skornyakov},\ and\ \citenamefont {Savrasov}}]{Leonov2020}%
  \BibitemOpen
  \bibfield  {author} {\bibinfo {author} {\bibfnamefont {I.}~\bibnamefont {Leonov}}, \bibinfo {author} {\bibfnamefont {S.~L.}\ \bibnamefont {Skornyakov}},\ and\ \bibinfo {author} {\bibfnamefont {S.~Y.}\ \bibnamefont {Savrasov}},\ }\bibfield  {title} {\bibinfo {title} {{Lifshitz transition and frustration of magnetic moments in infinite-layer ${\mathrm{NdNiO}}_{2}$ upon hole doping}},\ }\href {https://doi.org/10.1103/PhysRevB.101.241108} {\bibfield  {journal} {\bibinfo  {journal} {Phys. Rev. B}\ }\textbf {\bibinfo {volume} {101}},\ \bibinfo {pages} {241108} (\bibinfo {year} {2020})}\BibitemShut {NoStop}%
\bibitem [{\citenamefont {Liu}\ \emph {et~al.}(2020)\citenamefont {Liu}, \citenamefont {Ren}, \citenamefont {Zhu}, \citenamefont {Wang},\ and\ \citenamefont {Yang}}]{Liu2020}%
  \BibitemOpen
  \bibfield  {author} {\bibinfo {author} {\bibfnamefont {Z.}~\bibnamefont {Liu}}, \bibinfo {author} {\bibfnamefont {Z.}~\bibnamefont {Ren}}, \bibinfo {author} {\bibfnamefont {W.}~\bibnamefont {Zhu}}, \bibinfo {author} {\bibfnamefont {Z.}~\bibnamefont {Wang}},\ and\ \bibinfo {author} {\bibfnamefont {J.}~\bibnamefont {Yang}},\ }\bibfield  {title} {\bibinfo {title} {{Electronic and magnetic structure of infinite-layer NdNiO$_2$: trace of antiferromagnetic metal}},\ }\href {https://doi.org/10.1038/s41535-020-0229-1} {\bibfield  {journal} {\bibinfo  {journal} {npj Quantum Mater.}\ }\textbf {\bibinfo {volume} {5}},\ \bibinfo {pages} {31} (\bibinfo {year} {2020})}\BibitemShut {NoStop}%
\bibitem [{\citenamefont {Sakakibara}\ \emph {et~al.}(2020)\citenamefont {Sakakibara}, \citenamefont {Usui}, \citenamefont {Suzuki}, \citenamefont {Kotani}, \citenamefont {Aoki},\ and\ \citenamefont {Kuroki}}]{Sakakibara2020}%
  \BibitemOpen
  \bibfield  {author} {\bibinfo {author} {\bibfnamefont {H.}~\bibnamefont {Sakakibara}}, \bibinfo {author} {\bibfnamefont {H.}~\bibnamefont {Usui}}, \bibinfo {author} {\bibfnamefont {K.}~\bibnamefont {Suzuki}}, \bibinfo {author} {\bibfnamefont {T.}~\bibnamefont {Kotani}}, \bibinfo {author} {\bibfnamefont {H.}~\bibnamefont {Aoki}},\ and\ \bibinfo {author} {\bibfnamefont {K.}~\bibnamefont {Kuroki}},\ }\bibfield  {title} {\bibinfo {title} {{Model Construction and a Possibility of Cupratelike Pairing in a New ${d}^{9}$ Nickelate Superconductor $(\mathrm{Nd},\mathrm{Sr}){\mathrm{NiO}}_{2}$}},\ }\href {https://doi.org/10.1103/PhysRevLett.125.077003} {\bibfield  {journal} {\bibinfo  {journal} {Phys. Rev. Lett.}\ }\textbf {\bibinfo {volume} {125}},\ \bibinfo {pages} {077003} (\bibinfo {year} {2020})}\BibitemShut {NoStop}%
\bibitem [{\citenamefont {Wu}\ \emph {et~al.}(2020)\citenamefont {Wu}, \citenamefont {Di~Sante}, \citenamefont {Schwemmer}, \citenamefont {Hanke}, \citenamefont {Hwang}, \citenamefont {Raghu},\ and\ \citenamefont {Thomale}}]{Wu2020}%
  \BibitemOpen
  \bibfield  {author} {\bibinfo {author} {\bibfnamefont {X.}~\bibnamefont {Wu}}, \bibinfo {author} {\bibfnamefont {D.}~\bibnamefont {Di~Sante}}, \bibinfo {author} {\bibfnamefont {T.}~\bibnamefont {Schwemmer}}, \bibinfo {author} {\bibfnamefont {W.}~\bibnamefont {Hanke}}, \bibinfo {author} {\bibfnamefont {H.~Y.}\ \bibnamefont {Hwang}}, \bibinfo {author} {\bibfnamefont {S.}~\bibnamefont {Raghu}},\ and\ \bibinfo {author} {\bibfnamefont {R.}~\bibnamefont {Thomale}},\ }\bibfield  {title} {\bibinfo {title} {{Robust ${d}_{{x}^{2}\ensuremath{-}{y}^{2}}$-wave superconductivity of infinite-layer nickelates}},\ }\href {https://doi.org/10.1103/PhysRevB.101.060504} {\bibfield  {journal} {\bibinfo  {journal} {Phys. Rev. B}\ }\textbf {\bibinfo {volume} {101}},\ \bibinfo {pages} {060504} (\bibinfo {year} {2020})}\BibitemShut {NoStop}%
\bibitem [{\citenamefont {Been}\ \emph {et~al.}(2021)\citenamefont {Been}, \citenamefont {Lee}, \citenamefont {Hwang}, \citenamefont {Cui}, \citenamefont {Zaanen}, \citenamefont {Devereaux}, \citenamefont {Moritz},\ and\ \citenamefont {Jia}}]{Been2021}%
  \BibitemOpen
  \bibfield  {author} {\bibinfo {author} {\bibfnamefont {E.}~\bibnamefont {Been}}, \bibinfo {author} {\bibfnamefont {W.-S.}\ \bibnamefont {Lee}}, \bibinfo {author} {\bibfnamefont {H.~Y.}\ \bibnamefont {Hwang}}, \bibinfo {author} {\bibfnamefont {Y.}~\bibnamefont {Cui}}, \bibinfo {author} {\bibfnamefont {J.}~\bibnamefont {Zaanen}}, \bibinfo {author} {\bibfnamefont {T.}~\bibnamefont {Devereaux}}, \bibinfo {author} {\bibfnamefont {B.}~\bibnamefont {Moritz}},\ and\ \bibinfo {author} {\bibfnamefont {C.}~\bibnamefont {Jia}},\ }\bibfield  {title} {\bibinfo {title} {{Electronic structure trends across the rare-earth series in superconducting infinite-layer nickelates}},\ }\href {https://doi.org/10.1103/PhysRevX.11.011050} {\bibfield  {journal} {\bibinfo  {journal} {Phys. Rev. X}\ }\textbf {\bibinfo {volume} {11}},\ \bibinfo {pages} {011050} (\bibinfo {year} {2021})}\BibitemShut {NoStop}%
\bibitem [{\citenamefont {Torrance}\ \emph {et~al.}(1992)\citenamefont {Torrance}, \citenamefont {Lacorre}, \citenamefont {Nazzal}, \citenamefont {Ansaldo},\ and\ \citenamefont {Niedermayer}}]{Torrance1992}%
  \BibitemOpen
  \bibfield  {author} {\bibinfo {author} {\bibfnamefont {J.~B.}\ \bibnamefont {Torrance}}, \bibinfo {author} {\bibfnamefont {P.}~\bibnamefont {Lacorre}}, \bibinfo {author} {\bibfnamefont {A.~I.}\ \bibnamefont {Nazzal}}, \bibinfo {author} {\bibfnamefont {E.~J.}\ \bibnamefont {Ansaldo}},\ and\ \bibinfo {author} {\bibfnamefont {C.}~\bibnamefont {Niedermayer}},\ }\bibfield  {title} {\bibinfo {title} {{Systematic study of insulator-metal transitions in perovskites R${\mathrm{NiO}}_{3}$ (R=Pr,Nd,Sm,Eu) due to closing of charge-transfer gap}},\ }\href {https://doi.org/10.1103/PhysRevB.45.8209} {\bibfield  {journal} {\bibinfo  {journal} {Phys. Rev. B}\ }\textbf {\bibinfo {volume} {45}},\ \bibinfo {pages} {8209} (\bibinfo {year} {1992})}\BibitemShut {NoStop}%
\bibitem [{\citenamefont {Middey}\ \emph {et~al.}(2016)\citenamefont {Middey}, \citenamefont {Chakhalian}, \citenamefont {Mahadevan}, \citenamefont {Freeland}, \citenamefont {Millis},\ and\ \citenamefont {Sarma}}]{Middey2016}%
  \BibitemOpen
  \bibfield  {author} {\bibinfo {author} {\bibfnamefont {S.}~\bibnamefont {Middey}}, \bibinfo {author} {\bibfnamefont {J.}~\bibnamefont {Chakhalian}}, \bibinfo {author} {\bibfnamefont {P.}~\bibnamefont {Mahadevan}}, \bibinfo {author} {\bibfnamefont {J.}~\bibnamefont {Freeland}}, \bibinfo {author} {\bibfnamefont {A.}~\bibnamefont {Millis}},\ and\ \bibinfo {author} {\bibfnamefont {D.}~\bibnamefont {Sarma}},\ }\bibfield  {title} {\bibinfo {title} {Physics of ultrathin films and heterostructures of rare-earth nickelates},\ }\href {https://doi.org/10.1146/annurev-matsci-070115-032057} {\bibfield  {journal} {\bibinfo  {journal} {Annual Review of Materials Research}\ }\textbf {\bibinfo {volume} {46}},\ \bibinfo {pages} {305} (\bibinfo {year} {2016})}\BibitemShut {NoStop}%
\bibitem [{\citenamefont {Zhang}\ \emph {et~al.}(2023)\citenamefont {Zhang}, \citenamefont {Zhang}, \citenamefont {He}, \citenamefont {Wang},\ and\ \citenamefont {Ghosez}}]{Zhang2023}%
  \BibitemOpen
  \bibfield  {author} {\bibinfo {author} {\bibfnamefont {Y.}~\bibnamefont {Zhang}}, \bibinfo {author} {\bibfnamefont {J.}~\bibnamefont {Zhang}}, \bibinfo {author} {\bibfnamefont {X.}~\bibnamefont {He}}, \bibinfo {author} {\bibfnamefont {J.}~\bibnamefont {Wang}},\ and\ \bibinfo {author} {\bibfnamefont {P.}~\bibnamefont {Ghosez}},\ }\bibfield  {title} {\bibinfo {title} {{Rare-earth control of phase transitions in infinite-layer nickelates}},\ }\bibfield  {journal} {\bibinfo  {journal} {PNAS Nexus}\ }\textbf {\bibinfo {volume} {2}},\ \href {https://doi.org/10.1093/pnasnexus/pgad108} {10.1093/pnasnexus/pgad108} (\bibinfo {year} {2023}),\ \bibinfo {note} {pgad108},\ \Eprint {https://arxiv.org/abs/https://academic.oup.com/pnasnexus/article-pdf/2/5/pgad108/50633207/pgad108.pdf} {https://academic.oup.com/pnasnexus/article-pdf/2/5/pgad108/50633207/pgad108.pdf} \BibitemShut {NoStop}%
\bibitem [{\citenamefont {Subedi}(2023)}]{Subedi2023}%
  \BibitemOpen
  \bibfield  {author} {\bibinfo {author} {\bibfnamefont {A.}~\bibnamefont {Subedi}},\ }\bibfield  {title} {\bibinfo {title} {Possible structural quantum criticality tuned by rare-earth ion substitution in infinite-layer nickelates},\ }\href {https://doi.org/10.1103/PhysRevMaterials.7.024801} {\bibfield  {journal} {\bibinfo  {journal} {Phys. Rev. Mater.}\ }\textbf {\bibinfo {volume} {7}},\ \bibinfo {pages} {024801} (\bibinfo {year} {2023})}\BibitemShut {NoStop}%
\bibitem [{\citenamefont {Lu}\ \emph {et~al.}(2021)\citenamefont {Lu}, \citenamefont {Rossi}, \citenamefont {Nag}, \citenamefont {Osada}, \citenamefont {Li}, \citenamefont {Lee}, \citenamefont {Wang}, \citenamefont {Garcia-Fernandez}, \citenamefont {Agrestini}, \citenamefont {Shen}, \citenamefont {Been}, \citenamefont {Moritz}, \citenamefont {Devereaux}, \citenamefont {Zaanen}, \citenamefont {Hwang}, \citenamefont {Zhou},\ and\ \citenamefont {Lee}}]{Lu2021}%
  \BibitemOpen
  \bibfield  {author} {\bibinfo {author} {\bibfnamefont {H.}~\bibnamefont {Lu}}, \bibinfo {author} {\bibfnamefont {M.}~\bibnamefont {Rossi}}, \bibinfo {author} {\bibfnamefont {A.}~\bibnamefont {Nag}}, \bibinfo {author} {\bibfnamefont {M.}~\bibnamefont {Osada}}, \bibinfo {author} {\bibfnamefont {D.~F.}\ \bibnamefont {Li}}, \bibinfo {author} {\bibfnamefont {K.}~\bibnamefont {Lee}}, \bibinfo {author} {\bibfnamefont {B.~Y.}\ \bibnamefont {Wang}}, \bibinfo {author} {\bibfnamefont {M.}~\bibnamefont {Garcia-Fernandez}}, \bibinfo {author} {\bibfnamefont {S.}~\bibnamefont {Agrestini}}, \bibinfo {author} {\bibfnamefont {Z.~X.}\ \bibnamefont {Shen}}, \bibinfo {author} {\bibfnamefont {E.~M.}\ \bibnamefont {Been}}, \bibinfo {author} {\bibfnamefont {B.}~\bibnamefont {Moritz}}, \bibinfo {author} {\bibfnamefont {T.~P.}\ \bibnamefont {Devereaux}}, \bibinfo {author} {\bibfnamefont {J.}~\bibnamefont {Zaanen}}, \bibinfo {author} {\bibfnamefont {H.~Y.}\ \bibnamefont {Hwang}}, \bibinfo {author} {\bibfnamefont {K.-J.}\ \bibnamefont
  {Zhou}},\ and\ \bibinfo {author} {\bibfnamefont {W.~S.}\ \bibnamefont {Lee}},\ }\bibfield  {title} {\bibinfo {title} {Magnetic excitations in infinite-layer nickelates},\ }\href {https://doi.org/10.1126/science.abd7726} {\bibfield  {journal} {\bibinfo  {journal} {Science}\ }\textbf {\bibinfo {volume} {373}},\ \bibinfo {pages} {213} (\bibinfo {year} {2021})}\BibitemShut {NoStop}%
\bibitem [{\citenamefont {Gao}\ \emph {et~al.}(2022)\citenamefont {Gao}, \citenamefont {Fan}, \citenamefont {Wang}, \citenamefont {Li}, \citenamefont {Ren}, \citenamefont {Biało}, \citenamefont {Drewanowski}, \citenamefont {Rothenbühler}, \citenamefont {Choi}, \citenamefont {Wang}, \citenamefont {Xiang}, \citenamefont {Hu}, \citenamefont {Zhou}, \citenamefont {Bisogni}, \citenamefont {Comin}, \citenamefont {Chang}, \citenamefont {Pelliciari}, \citenamefont {Zhou},\ and\ \citenamefont {Zhu}}]{gao2022}%
  \BibitemOpen
  \bibfield  {author} {\bibinfo {author} {\bibfnamefont {Q.}~\bibnamefont {Gao}}, \bibinfo {author} {\bibfnamefont {S.}~\bibnamefont {Fan}}, \bibinfo {author} {\bibfnamefont {Q.}~\bibnamefont {Wang}}, \bibinfo {author} {\bibfnamefont {J.}~\bibnamefont {Li}}, \bibinfo {author} {\bibfnamefont {X.}~\bibnamefont {Ren}}, \bibinfo {author} {\bibfnamefont {I.}~\bibnamefont {Biało}}, \bibinfo {author} {\bibfnamefont {A.}~\bibnamefont {Drewanowski}}, \bibinfo {author} {\bibfnamefont {P.}~\bibnamefont {Rothenbühler}}, \bibinfo {author} {\bibfnamefont {J.}~\bibnamefont {Choi}}, \bibinfo {author} {\bibfnamefont {Y.}~\bibnamefont {Wang}}, \bibinfo {author} {\bibfnamefont {T.}~\bibnamefont {Xiang}}, \bibinfo {author} {\bibfnamefont {J.}~\bibnamefont {Hu}}, \bibinfo {author} {\bibfnamefont {K.-J.}\ \bibnamefont {Zhou}}, \bibinfo {author} {\bibfnamefont {V.}~\bibnamefont {Bisogni}}, \bibinfo {author} {\bibfnamefont {R.}~\bibnamefont {Comin}}, \bibinfo {author} {\bibfnamefont {J.}~\bibnamefont {Chang}}, \bibinfo {author}
  {\bibfnamefont {J.}~\bibnamefont {Pelliciari}}, \bibinfo {author} {\bibfnamefont {X.~J.}\ \bibnamefont {Zhou}},\ and\ \bibinfo {author} {\bibfnamefont {Z.}~\bibnamefont {Zhu}},\ }\href@noop {} {\bibinfo {title} {{Magnetic Excitations in Strained Infinite-layer Nickelate PrNiO$_2$}}} (\bibinfo {year} {2022}),\ \Eprint {https://arxiv.org/abs/2208.05614} {arXiv:2208.05614 [cond-mat.supr-con]} \BibitemShut {NoStop}%
\bibitem [{\citenamefont {Tam}\ \emph {et~al.}(2022)\citenamefont {Tam}, \citenamefont {Choi}, \citenamefont {Ding}, \citenamefont {Agrestini}, \citenamefont {Nag}, \citenamefont {Wu}, \citenamefont {Huang}, \citenamefont {Luo}, \citenamefont {Gao}, \citenamefont {Garc{\'i}a-Fern{\'a}ndez}, \citenamefont {Qiao},\ and\ \citenamefont {Zhou}}]{Tam2022}%
  \BibitemOpen
  \bibfield  {author} {\bibinfo {author} {\bibfnamefont {C.~C.}\ \bibnamefont {Tam}}, \bibinfo {author} {\bibfnamefont {J.}~\bibnamefont {Choi}}, \bibinfo {author} {\bibfnamefont {X.}~\bibnamefont {Ding}}, \bibinfo {author} {\bibfnamefont {S.}~\bibnamefont {Agrestini}}, \bibinfo {author} {\bibfnamefont {A.}~\bibnamefont {Nag}}, \bibinfo {author} {\bibfnamefont {M.}~\bibnamefont {Wu}}, \bibinfo {author} {\bibfnamefont {B.}~\bibnamefont {Huang}}, \bibinfo {author} {\bibfnamefont {H.}~\bibnamefont {Luo}}, \bibinfo {author} {\bibfnamefont {P.}~\bibnamefont {Gao}}, \bibinfo {author} {\bibfnamefont {M.}~\bibnamefont {Garc{\'i}a-Fern{\'a}ndez}}, \bibinfo {author} {\bibfnamefont {L.}~\bibnamefont {Qiao}},\ and\ \bibinfo {author} {\bibfnamefont {K.-J.}\ \bibnamefont {Zhou}},\ }\bibfield  {title} {\bibinfo {title} {{Charge density waves in infinite-layer NdNiO$_2$ nickelates}},\ }\href {https://doi.org/10.1038/s41563-022-01330-1} {\bibfield  {journal} {\bibinfo  {journal} {Nature Materials}\ }\textbf {\bibinfo {volume}
  {21}},\ \bibinfo {pages} {1116} (\bibinfo {year} {2022})}\BibitemShut {NoStop}%
\bibitem [{\citenamefont {Krieger}\ \emph {et~al.}(2022)\citenamefont {Krieger}, \citenamefont {Martinelli}, \citenamefont {Zeng}, \citenamefont {Chow}, \citenamefont {Kummer}, \citenamefont {Arpaia}, \citenamefont {Moretti~Sala}, \citenamefont {Brookes}, \citenamefont {Ariando}, \citenamefont {Viart}, \citenamefont {Salluzzo}, \citenamefont {Ghiringhelli},\ and\ \citenamefont {Preziosi}}]{Krieger2022}%
  \BibitemOpen
  \bibfield  {author} {\bibinfo {author} {\bibfnamefont {G.}~\bibnamefont {Krieger}}, \bibinfo {author} {\bibfnamefont {L.}~\bibnamefont {Martinelli}}, \bibinfo {author} {\bibfnamefont {S.}~\bibnamefont {Zeng}}, \bibinfo {author} {\bibfnamefont {L.~E.}\ \bibnamefont {Chow}}, \bibinfo {author} {\bibfnamefont {K.}~\bibnamefont {Kummer}}, \bibinfo {author} {\bibfnamefont {R.}~\bibnamefont {Arpaia}}, \bibinfo {author} {\bibfnamefont {M.}~\bibnamefont {Moretti~Sala}}, \bibinfo {author} {\bibfnamefont {N.~B.}\ \bibnamefont {Brookes}}, \bibinfo {author} {\bibfnamefont {A.}~\bibnamefont {Ariando}}, \bibinfo {author} {\bibfnamefont {N.}~\bibnamefont {Viart}}, \bibinfo {author} {\bibfnamefont {M.}~\bibnamefont {Salluzzo}}, \bibinfo {author} {\bibfnamefont {G.}~\bibnamefont {Ghiringhelli}},\ and\ \bibinfo {author} {\bibfnamefont {D.}~\bibnamefont {Preziosi}},\ }\bibfield  {title} {\bibinfo {title} {{Charge and Spin Order Dichotomy in ${\mathrm{NdNiO}}_{2}$ Driven by the Capping Layer}},\ }\href
  {https://doi.org/10.1103/PhysRevLett.129.027002} {\bibfield  {journal} {\bibinfo  {journal} {Phys. Rev. Lett.}\ }\textbf {\bibinfo {volume} {129}},\ \bibinfo {pages} {027002} (\bibinfo {year} {2022})}\BibitemShut {NoStop}%
\bibitem [{\citenamefont {Ren}\ \emph {et~al.}(2023)\citenamefont {Ren}, \citenamefont {Sutarto}, \citenamefont {Gao}, \citenamefont {Wang}, \citenamefont {Li}, \citenamefont {Wang}, \citenamefont {Xiang}, \citenamefont {Hu}, \citenamefont {Zhang}, \citenamefont {Chang}, \citenamefont {Comin}, \citenamefont {Zhou},\ and\ \citenamefont {Zhu}}]{Ren2023}%
  \BibitemOpen
  \bibfield  {author} {\bibinfo {author} {\bibfnamefont {X.}~\bibnamefont {Ren}}, \bibinfo {author} {\bibfnamefont {R.}~\bibnamefont {Sutarto}}, \bibinfo {author} {\bibfnamefont {Q.}~\bibnamefont {Gao}}, \bibinfo {author} {\bibfnamefont {Q.}~\bibnamefont {Wang}}, \bibinfo {author} {\bibfnamefont {J.}~\bibnamefont {Li}}, \bibinfo {author} {\bibfnamefont {Y.}~\bibnamefont {Wang}}, \bibinfo {author} {\bibfnamefont {T.}~\bibnamefont {Xiang}}, \bibinfo {author} {\bibfnamefont {J.}~\bibnamefont {Hu}}, \bibinfo {author} {\bibfnamefont {F.-C.}\ \bibnamefont {Zhang}}, \bibinfo {author} {\bibfnamefont {J.}~\bibnamefont {Chang}}, \bibinfo {author} {\bibfnamefont {R.}~\bibnamefont {Comin}}, \bibinfo {author} {\bibfnamefont {X.~J.}\ \bibnamefont {Zhou}},\ and\ \bibinfo {author} {\bibfnamefont {Z.}~\bibnamefont {Zhu}},\ }\href@noop {} {\bibinfo {title} {Symmetry of charge order in infinite-layer nickelates}} (\bibinfo {year} {2023}),\ \Eprint {https://arxiv.org/abs/2303.02865} {arXiv:2303.02865 [cond-mat.supr-con]}
  \BibitemShut {NoStop}%
\bibitem [{\citenamefont {Lee}\ \emph {et~al.}(2020)\citenamefont {Lee}, \citenamefont {Goodge}, \citenamefont {Li}, \citenamefont {Osada}, \citenamefont {Wang}, \citenamefont {Cui}, \citenamefont {Kourkoutis},\ and\ \citenamefont {Hwang}}]{Lee2020}%
  \BibitemOpen
  \bibfield  {author} {\bibinfo {author} {\bibfnamefont {K.}~\bibnamefont {Lee}}, \bibinfo {author} {\bibfnamefont {B.~H.}\ \bibnamefont {Goodge}}, \bibinfo {author} {\bibfnamefont {D.}~\bibnamefont {Li}}, \bibinfo {author} {\bibfnamefont {M.}~\bibnamefont {Osada}}, \bibinfo {author} {\bibfnamefont {B.~Y.}\ \bibnamefont {Wang}}, \bibinfo {author} {\bibfnamefont {Y.}~\bibnamefont {Cui}}, \bibinfo {author} {\bibfnamefont {L.~F.}\ \bibnamefont {Kourkoutis}},\ and\ \bibinfo {author} {\bibfnamefont {H.~Y.}\ \bibnamefont {Hwang}},\ }\bibfield  {title} {\bibinfo {title} {{Aspects of the synthesis of thin film superconducting infinite-layer nickelates}},\ }\href {https://doi.org/10.1063/5.0005103} {\bibfield  {journal} {\bibinfo  {journal} {APL Mater.}\ }\textbf {\bibinfo {volume} {8}},\ \bibinfo {pages} {041107} (\bibinfo {year} {2020})}\BibitemShut {NoStop}%
\bibitem [{\citenamefont {Osada}\ \emph {et~al.}(2020{\natexlab{b}})\citenamefont {Osada}, \citenamefont {Wang}, \citenamefont {Goodge}, \citenamefont {Lee}, \citenamefont {Yoon}, \citenamefont {Sakuma}, \citenamefont {Li}, \citenamefont {Miura}, \citenamefont {Kourkoutis},\ and\ \citenamefont {Hwang}}]{Osada2020b}%
  \BibitemOpen
  \bibfield  {author} {\bibinfo {author} {\bibfnamefont {M.}~\bibnamefont {Osada}}, \bibinfo {author} {\bibfnamefont {B.~Y.}\ \bibnamefont {Wang}}, \bibinfo {author} {\bibfnamefont {B.~H.}\ \bibnamefont {Goodge}}, \bibinfo {author} {\bibfnamefont {K.}~\bibnamefont {Lee}}, \bibinfo {author} {\bibfnamefont {H.}~\bibnamefont {Yoon}}, \bibinfo {author} {\bibfnamefont {K.}~\bibnamefont {Sakuma}}, \bibinfo {author} {\bibfnamefont {D.}~\bibnamefont {Li}}, \bibinfo {author} {\bibfnamefont {M.}~\bibnamefont {Miura}}, \bibinfo {author} {\bibfnamefont {L.~F.}\ \bibnamefont {Kourkoutis}},\ and\ \bibinfo {author} {\bibfnamefont {H.~Y.}\ \bibnamefont {Hwang}},\ }\bibfield  {title} {\bibinfo {title} {A superconducting praseodymium nickelate with infinite layer structure},\ }\href {https://doi.org/10.1021/acs.nanolett.0c01392} {\bibfield  {journal} {\bibinfo  {journal} {Nano Letters}\ }\textbf {\bibinfo {volume} {20}},\ \bibinfo {pages} {5735} (\bibinfo {year} {2020}{\natexlab{b}})},\ \bibinfo {note} {pMID: 32574061},\
  \Eprint {https://arxiv.org/abs/https://doi.org/10.1021/acs.nanolett.0c01392} {https://doi.org/10.1021/acs.nanolett.0c01392} \BibitemShut {NoStop}%
\bibitem [{\citenamefont {Raji}\ \emph {et~al.}(2023)\citenamefont {Raji}, \citenamefont {Krieger}, \citenamefont {Viart}, \citenamefont {Preziosi}, \citenamefont {Rueff},\ and\ \citenamefont {Gloter}}]{raji2023}%
  \BibitemOpen
  \bibfield  {author} {\bibinfo {author} {\bibfnamefont {A.}~\bibnamefont {Raji}}, \bibinfo {author} {\bibfnamefont {G.}~\bibnamefont {Krieger}}, \bibinfo {author} {\bibfnamefont {N.}~\bibnamefont {Viart}}, \bibinfo {author} {\bibfnamefont {D.}~\bibnamefont {Preziosi}}, \bibinfo {author} {\bibfnamefont {J.-P.}\ \bibnamefont {Rueff}},\ and\ \bibinfo {author} {\bibfnamefont {A.}~\bibnamefont {Gloter}},\ }\href@noop {} {\bibinfo {title} {Charge distribution across capped and uncapped infinite-layer neodymium nickelate thin films}} (\bibinfo {year} {2023}),\ \Eprint {https://arxiv.org/abs/2306.10507} {arXiv:2306.10507 [cond-mat.mtrl-sci]} \BibitemShut {NoStop}%
\bibitem [{\citenamefont {Parzyck}\ \emph {et~al.}(2023)\citenamefont {Parzyck}, \citenamefont {Gupta}, \citenamefont {Wu}, \citenamefont {Anil}, \citenamefont {Bhatt}, \citenamefont {Bouliane}, \citenamefont {Gong}, \citenamefont {Gregory}, \citenamefont {Luo}, \citenamefont {Sutarto}, \citenamefont {He}, \citenamefont {Chuang}, \citenamefont {Zhou}, \citenamefont {Herranz}, \citenamefont {Kourkoutis}, \citenamefont {Singer}, \citenamefont {Schlom}, \citenamefont {Hawthorn},\ and\ \citenamefont {Shen}}]{parzyck2023}%
  \BibitemOpen
  \bibfield  {author} {\bibinfo {author} {\bibfnamefont {C.~T.}\ \bibnamefont {Parzyck}}, \bibinfo {author} {\bibfnamefont {N.~K.}\ \bibnamefont {Gupta}}, \bibinfo {author} {\bibfnamefont {Y.}~\bibnamefont {Wu}}, \bibinfo {author} {\bibfnamefont {V.}~\bibnamefont {Anil}}, \bibinfo {author} {\bibfnamefont {L.}~\bibnamefont {Bhatt}}, \bibinfo {author} {\bibfnamefont {M.}~\bibnamefont {Bouliane}}, \bibinfo {author} {\bibfnamefont {R.}~\bibnamefont {Gong}}, \bibinfo {author} {\bibfnamefont {B.~Z.}\ \bibnamefont {Gregory}}, \bibinfo {author} {\bibfnamefont {A.}~\bibnamefont {Luo}}, \bibinfo {author} {\bibfnamefont {R.}~\bibnamefont {Sutarto}}, \bibinfo {author} {\bibfnamefont {F.}~\bibnamefont {He}}, \bibinfo {author} {\bibfnamefont {Y.~D.}\ \bibnamefont {Chuang}}, \bibinfo {author} {\bibfnamefont {T.}~\bibnamefont {Zhou}}, \bibinfo {author} {\bibfnamefont {G.}~\bibnamefont {Herranz}}, \bibinfo {author} {\bibfnamefont {L.~F.}\ \bibnamefont {Kourkoutis}}, \bibinfo {author} {\bibfnamefont {A.}~\bibnamefont
  {Singer}}, \bibinfo {author} {\bibfnamefont {D.~G.}\ \bibnamefont {Schlom}}, \bibinfo {author} {\bibfnamefont {D.~G.}\ \bibnamefont {Hawthorn}},\ and\ \bibinfo {author} {\bibfnamefont {K.~M.}\ \bibnamefont {Shen}},\ }\href@noop {} {\bibinfo {title} {Absence of $3a_0$ charge density wave order in the infinite layer nickelates}} (\bibinfo {year} {2023}),\ \Eprint {https://arxiv.org/abs/2307.06486} {arXiv:2307.06486 [cond-mat.supr-con]} \BibitemShut {NoStop}%
\bibitem [{SM()}]{SM}%
  \BibitemOpen
  \href@noop {} {}\bibinfo {note} {See Supplemental Material at [URL] for additional experiment details, data and analysis.}\BibitemShut {Stop}%
\bibitem [{\citenamefont {Sala}\ \emph {et~al.}(2011)\citenamefont {Sala}, \citenamefont {Bisogni}, \citenamefont {Aruta}, \citenamefont {Balestrino}, \citenamefont {Berger}, \citenamefont {Brookes}, \citenamefont {de~Luca}, \citenamefont {Castro}, \citenamefont {Grioni}, \citenamefont {Guarise}, \citenamefont {Medaglia}, \citenamefont {Granozio}, \citenamefont {Minola}, \citenamefont {Perna}, \citenamefont {Radovic}, \citenamefont {Salluzzo}, \citenamefont {Schmitt}, \citenamefont {Zhou}, \citenamefont {Braicovich},\ and\ \citenamefont {Ghiringhelli}}]{Moretti2011}%
  \BibitemOpen
  \bibfield  {author} {\bibinfo {author} {\bibfnamefont {M.~M.}\ \bibnamefont {Sala}}, \bibinfo {author} {\bibfnamefont {V.}~\bibnamefont {Bisogni}}, \bibinfo {author} {\bibfnamefont {C.}~\bibnamefont {Aruta}}, \bibinfo {author} {\bibfnamefont {G.}~\bibnamefont {Balestrino}}, \bibinfo {author} {\bibfnamefont {H.}~\bibnamefont {Berger}}, \bibinfo {author} {\bibfnamefont {N.~B.}\ \bibnamefont {Brookes}}, \bibinfo {author} {\bibfnamefont {G.~M.}\ \bibnamefont {de~Luca}}, \bibinfo {author} {\bibfnamefont {D.~D.}\ \bibnamefont {Castro}}, \bibinfo {author} {\bibfnamefont {M.}~\bibnamefont {Grioni}}, \bibinfo {author} {\bibfnamefont {M.}~\bibnamefont {Guarise}}, \bibinfo {author} {\bibfnamefont {P.~G.}\ \bibnamefont {Medaglia}}, \bibinfo {author} {\bibfnamefont {F.~M.}\ \bibnamefont {Granozio}}, \bibinfo {author} {\bibfnamefont {M.}~\bibnamefont {Minola}}, \bibinfo {author} {\bibfnamefont {P.}~\bibnamefont {Perna}}, \bibinfo {author} {\bibfnamefont {M.}~\bibnamefont {Radovic}}, \bibinfo {author} {\bibfnamefont
  {M.}~\bibnamefont {Salluzzo}}, \bibinfo {author} {\bibfnamefont {T.}~\bibnamefont {Schmitt}}, \bibinfo {author} {\bibfnamefont {K.~J.}\ \bibnamefont {Zhou}}, \bibinfo {author} {\bibfnamefont {L.}~\bibnamefont {Braicovich}},\ and\ \bibinfo {author} {\bibfnamefont {G.}~\bibnamefont {Ghiringhelli}},\ }\bibfield  {title} {\bibinfo {title} {{Energy and symmetry of dd excitations in undoped layered cuprates measured by Cu $L_3$ resonant inelastic x-ray scattering}},\ }\href {https://doi.org/10.1088/1367-2630/13/4/043026} {\bibfield  {journal} {\bibinfo  {journal} {New J. Phys.}\ }\textbf {\bibinfo {volume} {13}},\ \bibinfo {pages} {043026} (\bibinfo {year} {2011})}\BibitemShut {NoStop}%
\bibitem [{\citenamefont {Lamsal}\ and\ \citenamefont {Montfrooij}(2016)}]{Lamsal2016}%
  \BibitemOpen
  \bibfield  {author} {\bibinfo {author} {\bibfnamefont {J.}~\bibnamefont {Lamsal}}\ and\ \bibinfo {author} {\bibfnamefont {W.}~\bibnamefont {Montfrooij}},\ }\bibfield  {title} {\bibinfo {title} {Extracting paramagnon excitations from resonant inelastic x-ray scattering experiments},\ }\href {https://doi.org/10.1103/PhysRevB.93.214513} {\bibfield  {journal} {\bibinfo  {journal} {Phys. Rev. B}\ }\textbf {\bibinfo {volume} {93}},\ \bibinfo {pages} {214513} (\bibinfo {year} {2016})}\BibitemShut {NoStop}%
\bibitem [{\citenamefont {Coldea}\ \emph {et~al.}(2001)\citenamefont {Coldea}, \citenamefont {Hayden}, \citenamefont {Aeppli}, \citenamefont {Perring}, \citenamefont {Frost}, \citenamefont {Mason}, \citenamefont {Cheong},\ and\ \citenamefont {Fisk}}]{Coldea2001}%
  \BibitemOpen
  \bibfield  {author} {\bibinfo {author} {\bibfnamefont {R.}~\bibnamefont {Coldea}}, \bibinfo {author} {\bibfnamefont {S.~M.}\ \bibnamefont {Hayden}}, \bibinfo {author} {\bibfnamefont {G.}~\bibnamefont {Aeppli}}, \bibinfo {author} {\bibfnamefont {T.~G.}\ \bibnamefont {Perring}}, \bibinfo {author} {\bibfnamefont {C.~D.}\ \bibnamefont {Frost}}, \bibinfo {author} {\bibfnamefont {T.~E.}\ \bibnamefont {Mason}}, \bibinfo {author} {\bibfnamefont {S.-W.}\ \bibnamefont {Cheong}},\ and\ \bibinfo {author} {\bibfnamefont {Z.}~\bibnamefont {Fisk}},\ }\bibfield  {title} {\bibinfo {title} {{Spin Waves and Electronic Interactions in ${\mathrm{La}}_{2}{\mathrm{CuO}}_{4}$}},\ }\href {https://doi.org/10.1103/PhysRevLett.86.5377} {\bibfield  {journal} {\bibinfo  {journal} {Phys. Rev. Lett.}\ }\textbf {\bibinfo {volume} {86}},\ \bibinfo {pages} {5377} (\bibinfo {year} {2001})}\BibitemShut {NoStop}%
\bibitem [{\citenamefont {Hayward}\ and\ \citenamefont {Rosseinsky}(2003)}]{Hayward2003}%
  \BibitemOpen
  \bibfield  {author} {\bibinfo {author} {\bibfnamefont {M.}~\bibnamefont {Hayward}}\ and\ \bibinfo {author} {\bibfnamefont {M.}~\bibnamefont {Rosseinsky}},\ }\bibfield  {title} {\bibinfo {title} {{Synthesis of the infinite layer Ni(I) phase NdNiO$_{2+x}$ by low temperature reduction of NdNiO$_3$ with sodium hydride}},\ }\href {https://doi.org/https://doi.org/10.1016/S1293-2558(03)00111-0} {\bibfield  {journal} {\bibinfo  {journal} {Solid State Sci.}\ }\textbf {\bibinfo {volume} {5}},\ \bibinfo {pages} {839 } (\bibinfo {year} {2003})}\BibitemShut {NoStop}%
\bibitem [{\citenamefont {Lin}\ \emph {et~al.}(2022)\citenamefont {Lin}, \citenamefont {Gawryluk}, \citenamefont {Klein}, \citenamefont {Huangfu}, \citenamefont {Pomjakushina}, \citenamefont {von Rohr},\ and\ \citenamefont {Schilling}}]{Lin_2022}%
  \BibitemOpen
  \bibfield  {author} {\bibinfo {author} {\bibfnamefont {H.}~\bibnamefont {Lin}}, \bibinfo {author} {\bibfnamefont {D.~J.}\ \bibnamefont {Gawryluk}}, \bibinfo {author} {\bibfnamefont {Y.~M.}\ \bibnamefont {Klein}}, \bibinfo {author} {\bibfnamefont {S.}~\bibnamefont {Huangfu}}, \bibinfo {author} {\bibfnamefont {E.}~\bibnamefont {Pomjakushina}}, \bibinfo {author} {\bibfnamefont {F.}~\bibnamefont {von Rohr}},\ and\ \bibinfo {author} {\bibfnamefont {A.}~\bibnamefont {Schilling}},\ }\bibfield  {title} {\bibinfo {title} {{Universal spin-glass behaviour in bulk LaNiO$_2$, PrNiO$_2$ and NdNiO$_2$}},\ }\href {https://doi.org/10.1088/1367-2630/ac465e} {\bibfield  {journal} {\bibinfo  {journal} {New Journal of Physics}\ }\textbf {\bibinfo {volume} {24}},\ \bibinfo {pages} {013022} (\bibinfo {year} {2022})}\BibitemShut {NoStop}%
\bibitem [{\citenamefont {Ortiz}\ \emph {et~al.}(2022)\citenamefont {Ortiz}, \citenamefont {Puphal}, \citenamefont {Klett}, \citenamefont {Hotz}, \citenamefont {Kremer}, \citenamefont {Trepka}, \citenamefont {Hemmida}, \citenamefont {von Nidda}, \citenamefont {Isobe}, \citenamefont {Khasanov}, \citenamefont {Luetkens}, \citenamefont {Hansmann}, \citenamefont {Keimer}, \citenamefont {Sch\"afer},\ and\ \citenamefont {Hepting}}]{Ortiz2022}%
  \BibitemOpen
  \bibfield  {author} {\bibinfo {author} {\bibfnamefont {R.~A.}\ \bibnamefont {Ortiz}}, \bibinfo {author} {\bibfnamefont {P.}~\bibnamefont {Puphal}}, \bibinfo {author} {\bibfnamefont {M.}~\bibnamefont {Klett}}, \bibinfo {author} {\bibfnamefont {F.}~\bibnamefont {Hotz}}, \bibinfo {author} {\bibfnamefont {R.~K.}\ \bibnamefont {Kremer}}, \bibinfo {author} {\bibfnamefont {H.}~\bibnamefont {Trepka}}, \bibinfo {author} {\bibfnamefont {M.}~\bibnamefont {Hemmida}}, \bibinfo {author} {\bibfnamefont {H.-A.~K.}\ \bibnamefont {von Nidda}}, \bibinfo {author} {\bibfnamefont {M.}~\bibnamefont {Isobe}}, \bibinfo {author} {\bibfnamefont {R.}~\bibnamefont {Khasanov}}, \bibinfo {author} {\bibfnamefont {H.}~\bibnamefont {Luetkens}}, \bibinfo {author} {\bibfnamefont {P.}~\bibnamefont {Hansmann}}, \bibinfo {author} {\bibfnamefont {B.}~\bibnamefont {Keimer}}, \bibinfo {author} {\bibfnamefont {T.}~\bibnamefont {Sch\"afer}},\ and\ \bibinfo {author} {\bibfnamefont {M.}~\bibnamefont {Hepting}},\ }\bibfield  {title} {\bibinfo {title}
  {Magnetic correlations in infinite-layer nickelates: An experimental and theoretical multimethod study},\ }\href {https://doi.org/10.1103/PhysRevResearch.4.023093} {\bibfield  {journal} {\bibinfo  {journal} {Phys. Rev. Res.}\ }\textbf {\bibinfo {volume} {4}},\ \bibinfo {pages} {023093} (\bibinfo {year} {2022})}\BibitemShut {NoStop}%
\bibitem [{\citenamefont {Ghiringhelli}\ \emph {et~al.}(2012)\citenamefont {Ghiringhelli}, \citenamefont {Tacon}, \citenamefont {Minola}, \citenamefont {Blanco-Canosa}, \citenamefont {Mazzoli}, \citenamefont {Brookes}, \citenamefont {Luca}, \citenamefont {Frano}, \citenamefont {Hawthorn}, \citenamefont {He}, \citenamefont {Loew}, \citenamefont {Sala}, \citenamefont {Peets}, \citenamefont {Salluzzo}, \citenamefont {Schierle}, \citenamefont {Sutarto}, \citenamefont {Sawatzky}, \citenamefont {Weschke}, \citenamefont {Keimer},\ and\ \citenamefont {Braicovich}}]{Ghiringhelli2012YBCO}%
  \BibitemOpen
  \bibfield  {author} {\bibinfo {author} {\bibfnamefont {G.}~\bibnamefont {Ghiringhelli}}, \bibinfo {author} {\bibfnamefont {M.~L.}\ \bibnamefont {Tacon}}, \bibinfo {author} {\bibfnamefont {M.}~\bibnamefont {Minola}}, \bibinfo {author} {\bibfnamefont {S.}~\bibnamefont {Blanco-Canosa}}, \bibinfo {author} {\bibfnamefont {C.}~\bibnamefont {Mazzoli}}, \bibinfo {author} {\bibfnamefont {N.~B.}\ \bibnamefont {Brookes}}, \bibinfo {author} {\bibfnamefont {G.~M.~D.}\ \bibnamefont {Luca}}, \bibinfo {author} {\bibfnamefont {A.}~\bibnamefont {Frano}}, \bibinfo {author} {\bibfnamefont {D.~G.}\ \bibnamefont {Hawthorn}}, \bibinfo {author} {\bibfnamefont {F.}~\bibnamefont {He}}, \bibinfo {author} {\bibfnamefont {T.}~\bibnamefont {Loew}}, \bibinfo {author} {\bibfnamefont {M.~M.}\ \bibnamefont {Sala}}, \bibinfo {author} {\bibfnamefont {D.~C.}\ \bibnamefont {Peets}}, \bibinfo {author} {\bibfnamefont {M.}~\bibnamefont {Salluzzo}}, \bibinfo {author} {\bibfnamefont {E.}~\bibnamefont {Schierle}}, \bibinfo {author} {\bibfnamefont
  {R.}~\bibnamefont {Sutarto}}, \bibinfo {author} {\bibfnamefont {G.~A.}\ \bibnamefont {Sawatzky}}, \bibinfo {author} {\bibfnamefont {E.}~\bibnamefont {Weschke}}, \bibinfo {author} {\bibfnamefont {B.}~\bibnamefont {Keimer}},\ and\ \bibinfo {author} {\bibfnamefont {L.}~\bibnamefont {Braicovich}},\ }\bibfield  {title} {\bibinfo {title} {{Long-Range Incommensurate Charge Fluctuations in (Y, Nd)Ba$_2$Cu$_3$O$_{6+x}$}},\ }\href {https://doi.org/10.1126/science.1223532} {\bibfield  {journal} {\bibinfo  {journal} {Science}\ }\textbf {\bibinfo {volume} {337}},\ \bibinfo {pages} {821} (\bibinfo {year} {2012})}\BibitemShut {NoStop}%
\bibitem [{\citenamefont {Pelliciari}\ \emph {et~al.}(2023)\citenamefont {Pelliciari}, \citenamefont {Khan}, \citenamefont {Wasik}, \citenamefont {Barbour}, \citenamefont {Li}, \citenamefont {Nie}, \citenamefont {Tranquada}, \citenamefont {Bisogni},\ and\ \citenamefont {Mazzoli}}]{pelliciari2023comment}%
  \BibitemOpen
  \bibfield  {author} {\bibinfo {author} {\bibfnamefont {J.}~\bibnamefont {Pelliciari}}, \bibinfo {author} {\bibfnamefont {N.}~\bibnamefont {Khan}}, \bibinfo {author} {\bibfnamefont {P.}~\bibnamefont {Wasik}}, \bibinfo {author} {\bibfnamefont {A.}~\bibnamefont {Barbour}}, \bibinfo {author} {\bibfnamefont {Y.}~\bibnamefont {Li}}, \bibinfo {author} {\bibfnamefont {Y.}~\bibnamefont {Nie}}, \bibinfo {author} {\bibfnamefont {J.~M.}\ \bibnamefont {Tranquada}}, \bibinfo {author} {\bibfnamefont {V.}~\bibnamefont {Bisogni}},\ and\ \bibinfo {author} {\bibfnamefont {C.}~\bibnamefont {Mazzoli}},\ }\href@noop {} {\bibinfo {title} {Comment on newly found charge density waves in infinite layer nickelates}} (\bibinfo {year} {2023}),\ \Eprint {https://arxiv.org/abs/2306.15086} {arXiv:2306.15086 [cond-mat.supr-con]} \BibitemShut {NoStop}%
\bibitem [{\citenamefont {Tam}\ \emph {et~al.}(2023)\citenamefont {Tam}, \citenamefont {Choi}, \citenamefont {Ding}, \citenamefont {Agrestini}, \citenamefont {Nag}, \citenamefont {Wu}, \citenamefont {Huang}, \citenamefont {Luo}, \citenamefont {Gao}, \citenamefont {Garcia-Fernandez}, \citenamefont {Qiao},\ and\ \citenamefont {Zhou}}]{tam2023reply}%
  \BibitemOpen
  \bibfield  {author} {\bibinfo {author} {\bibfnamefont {C.~C.}\ \bibnamefont {Tam}}, \bibinfo {author} {\bibfnamefont {J.}~\bibnamefont {Choi}}, \bibinfo {author} {\bibfnamefont {X.}~\bibnamefont {Ding}}, \bibinfo {author} {\bibfnamefont {S.}~\bibnamefont {Agrestini}}, \bibinfo {author} {\bibfnamefont {A.}~\bibnamefont {Nag}}, \bibinfo {author} {\bibfnamefont {M.}~\bibnamefont {Wu}}, \bibinfo {author} {\bibfnamefont {B.}~\bibnamefont {Huang}}, \bibinfo {author} {\bibfnamefont {H.}~\bibnamefont {Luo}}, \bibinfo {author} {\bibfnamefont {P.}~\bibnamefont {Gao}}, \bibinfo {author} {\bibfnamefont {M.}~\bibnamefont {Garcia-Fernandez}}, \bibinfo {author} {\bibfnamefont {L.}~\bibnamefont {Qiao}},\ and\ \bibinfo {author} {\bibfnamefont {K.-J.}\ \bibnamefont {Zhou}},\ }\href@noop {} {\bibinfo {title} {Reply to "comment on newly found charge density waves in infinite layer nickelates''}} (\bibinfo {year} {2023}),\ \Eprint {https://arxiv.org/abs/2307.13569} {arXiv:2307.13569 [cond-mat.str-el]} \BibitemShut {NoStop}%
\bibitem [{\citenamefont {Zeng}\ \emph {et~al.}(2022)\citenamefont {Zeng}, \citenamefont {Li}, \citenamefont {Chow}, \citenamefont {Cao}, \citenamefont {Zhang}, \citenamefont {Tang}, \citenamefont {Yin}, \citenamefont {Lim}, \citenamefont {Hu}, \citenamefont {Yang},\ and\ \citenamefont {Ariando}}]{Zeng2022}%
  \BibitemOpen
  \bibfield  {author} {\bibinfo {author} {\bibfnamefont {S.}~\bibnamefont {Zeng}}, \bibinfo {author} {\bibfnamefont {C.}~\bibnamefont {Li}}, \bibinfo {author} {\bibfnamefont {L.~E.}\ \bibnamefont {Chow}}, \bibinfo {author} {\bibfnamefont {Y.}~\bibnamefont {Cao}}, \bibinfo {author} {\bibfnamefont {Z.}~\bibnamefont {Zhang}}, \bibinfo {author} {\bibfnamefont {C.~S.}\ \bibnamefont {Tang}}, \bibinfo {author} {\bibfnamefont {X.}~\bibnamefont {Yin}}, \bibinfo {author} {\bibfnamefont {Z.~S.}\ \bibnamefont {Lim}}, \bibinfo {author} {\bibfnamefont {J.}~\bibnamefont {Hu}}, \bibinfo {author} {\bibfnamefont {P.}~\bibnamefont {Yang}},\ and\ \bibinfo {author} {\bibfnamefont {A.}~\bibnamefont {Ariando}},\ }\bibfield  {title} {\bibinfo {title} {{Superconductivity in infinite-layer nickelate La$_{1-x}$Ca$_x$NiO$_2$ thin films}},\ }\href {https://doi.org/10.1126/sciadv.abl9927} {\bibfield  {journal} {\bibinfo  {journal} {Science Advances}\ }\textbf {\bibinfo {volume} {8}},\ \bibinfo {pages} {eabl9927} (\bibinfo {year}
  {2022})},\ \Eprint {https://arxiv.org/abs/https://www.science.org/doi/pdf/10.1126/sciadv.abl9927} {https://www.science.org/doi/pdf/10.1126/sciadv.abl9927} \BibitemShut {NoStop}%
\bibitem [{\citenamefont {Peng}\ \emph {et~al.}(2022)\citenamefont {Peng}, \citenamefont {Jiang}, \citenamefont {Moritz}, \citenamefont {Devereaux},\ and\ \citenamefont {Jia}}]{peng2022}%
  \BibitemOpen
  \bibfield  {author} {\bibinfo {author} {\bibfnamefont {C.}~\bibnamefont {Peng}}, \bibinfo {author} {\bibfnamefont {H.-C.}\ \bibnamefont {Jiang}}, \bibinfo {author} {\bibfnamefont {B.}~\bibnamefont {Moritz}}, \bibinfo {author} {\bibfnamefont {T.~P.}\ \bibnamefont {Devereaux}},\ and\ \bibinfo {author} {\bibfnamefont {C.}~\bibnamefont {Jia}},\ }\href@noop {} {\bibinfo {title} {Charge order and superconductivity in a minimal two-band model for infinite-layer nickelates}} (\bibinfo {year} {2022}),\ \Eprint {https://arxiv.org/abs/2110.07593} {arXiv:2110.07593 [cond-mat.str-el]} \BibitemShut {NoStop}%
\bibitem [{\citenamefont {Kim}\ \emph {et~al.}(2018)\citenamefont {Kim}, \citenamefont {Souliou}, \citenamefont {Barber}, \citenamefont {Lefrançois}, \citenamefont {Minola}, \citenamefont {Tortora}, \citenamefont {Heid}, \citenamefont {Nandi}, \citenamefont {Borzi}, \citenamefont {Garbarino}, \citenamefont {Bosak}, \citenamefont {Porras}, \citenamefont {Loew}, \citenamefont {König}, \citenamefont {Moll}, \citenamefont {Mackenzie}, \citenamefont {Keimer}, \citenamefont {Hicks},\ and\ \citenamefont {Tacon}}]{Kim2018}%
  \BibitemOpen
  \bibfield  {author} {\bibinfo {author} {\bibfnamefont {H.-H.}\ \bibnamefont {Kim}}, \bibinfo {author} {\bibfnamefont {S.~M.}\ \bibnamefont {Souliou}}, \bibinfo {author} {\bibfnamefont {M.~E.}\ \bibnamefont {Barber}}, \bibinfo {author} {\bibfnamefont {E.}~\bibnamefont {Lefrançois}}, \bibinfo {author} {\bibfnamefont {M.}~\bibnamefont {Minola}}, \bibinfo {author} {\bibfnamefont {M.}~\bibnamefont {Tortora}}, \bibinfo {author} {\bibfnamefont {R.}~\bibnamefont {Heid}}, \bibinfo {author} {\bibfnamefont {N.}~\bibnamefont {Nandi}}, \bibinfo {author} {\bibfnamefont {R.~A.}\ \bibnamefont {Borzi}}, \bibinfo {author} {\bibfnamefont {G.}~\bibnamefont {Garbarino}}, \bibinfo {author} {\bibfnamefont {A.}~\bibnamefont {Bosak}}, \bibinfo {author} {\bibfnamefont {J.}~\bibnamefont {Porras}}, \bibinfo {author} {\bibfnamefont {T.}~\bibnamefont {Loew}}, \bibinfo {author} {\bibfnamefont {M.}~\bibnamefont {König}}, \bibinfo {author} {\bibfnamefont {P.~J.~W.}\ \bibnamefont {Moll}}, \bibinfo {author} {\bibfnamefont {A.~P.}\
  \bibnamefont {Mackenzie}}, \bibinfo {author} {\bibfnamefont {B.}~\bibnamefont {Keimer}}, \bibinfo {author} {\bibfnamefont {C.~W.}\ \bibnamefont {Hicks}},\ and\ \bibinfo {author} {\bibfnamefont {M.~L.}\ \bibnamefont {Tacon}},\ }\bibfield  {title} {\bibinfo {title} {Uniaxial pressure control of competing orders in a high-temperature superconductor},\ }\href {https://doi.org/10.1126/science.aat4708} {\bibfield  {journal} {\bibinfo  {journal} {Science}\ }\textbf {\bibinfo {volume} {362}},\ \bibinfo {pages} {1040} (\bibinfo {year} {2018})},\ \Eprint {https://arxiv.org/abs/https://www.science.org/doi/pdf/10.1126/science.aat4708} {https://www.science.org/doi/pdf/10.1126/science.aat4708} \BibitemShut {NoStop}%
\bibitem [{\citenamefont {Bluschke}\ \emph {et~al.}(2018)\citenamefont {Bluschke}, \citenamefont {Frano}, \citenamefont {Schierle}, \citenamefont {Putzky}, \citenamefont {Ghorbani}, \citenamefont {Ortiz}, \citenamefont {Suzuki}, \citenamefont {Christiani}, \citenamefont {Logvenov}, \citenamefont {Weschke}, \citenamefont {Birgeneau}, \citenamefont {da~Silva~Neto}, \citenamefont {Minola}, \citenamefont {Blanco-Canosa},\ and\ \citenamefont {Keimer}}]{Bluschke2018j}%
  \BibitemOpen
  \bibfield  {author} {\bibinfo {author} {\bibfnamefont {M.}~\bibnamefont {Bluschke}}, \bibinfo {author} {\bibfnamefont {A.}~\bibnamefont {Frano}}, \bibinfo {author} {\bibfnamefont {E.}~\bibnamefont {Schierle}}, \bibinfo {author} {\bibfnamefont {D.}~\bibnamefont {Putzky}}, \bibinfo {author} {\bibfnamefont {F.}~\bibnamefont {Ghorbani}}, \bibinfo {author} {\bibfnamefont {R.}~\bibnamefont {Ortiz}}, \bibinfo {author} {\bibfnamefont {H.}~\bibnamefont {Suzuki}}, \bibinfo {author} {\bibfnamefont {G.}~\bibnamefont {Christiani}}, \bibinfo {author} {\bibfnamefont {G.}~\bibnamefont {Logvenov}}, \bibinfo {author} {\bibfnamefont {E.}~\bibnamefont {Weschke}}, \bibinfo {author} {\bibfnamefont {R.~J.}\ \bibnamefont {Birgeneau}}, \bibinfo {author} {\bibfnamefont {E.~H.}\ \bibnamefont {da~Silva~Neto}}, \bibinfo {author} {\bibfnamefont {M.}~\bibnamefont {Minola}}, \bibinfo {author} {\bibfnamefont {S.}~\bibnamefont {Blanco-Canosa}},\ and\ \bibinfo {author} {\bibfnamefont {B.}~\bibnamefont {Keimer}},\ }\bibfield  {title}
  {\bibinfo {title} {Stabilization of three-dimensional charge order in {YBa$_2$Cu$_3$O$_{6+x}$} via epitaxial growth},\ }\href@noop {} {\bibfield  {journal} {\bibinfo  {journal} {Nature Communications}\ }\textbf {\bibinfo {volume} {9}},\ \bibinfo {pages} {2978} (\bibinfo {year} {2018})}\BibitemShut {NoStop}%
\bibitem [{\citenamefont {Choi}\ \emph {et~al.}(2022)\citenamefont {Choi}, \citenamefont {Wang}, \citenamefont {J\"ohr}, \citenamefont {Christensen}, \citenamefont {K\"uspert}, \citenamefont {Bucher}, \citenamefont {Biscette}, \citenamefont {Fischer}, \citenamefont {H\"ucker}, \citenamefont {Kurosawa}, \citenamefont {Momono}, \citenamefont {Oda}, \citenamefont {Ivashko}, \citenamefont {Zimmermann}, \citenamefont {Janoschek},\ and\ \citenamefont {Chang}}]{Choi2022}%
  \BibitemOpen
  \bibfield  {author} {\bibinfo {author} {\bibfnamefont {J.}~\bibnamefont {Choi}}, \bibinfo {author} {\bibfnamefont {Q.}~\bibnamefont {Wang}}, \bibinfo {author} {\bibfnamefont {S.}~\bibnamefont {J\"ohr}}, \bibinfo {author} {\bibfnamefont {N.~B.}\ \bibnamefont {Christensen}}, \bibinfo {author} {\bibfnamefont {J.}~\bibnamefont {K\"uspert}}, \bibinfo {author} {\bibfnamefont {D.}~\bibnamefont {Bucher}}, \bibinfo {author} {\bibfnamefont {D.}~\bibnamefont {Biscette}}, \bibinfo {author} {\bibfnamefont {M.~H.}\ \bibnamefont {Fischer}}, \bibinfo {author} {\bibfnamefont {M.}~\bibnamefont {H\"ucker}}, \bibinfo {author} {\bibfnamefont {T.}~\bibnamefont {Kurosawa}}, \bibinfo {author} {\bibfnamefont {N.}~\bibnamefont {Momono}}, \bibinfo {author} {\bibfnamefont {M.}~\bibnamefont {Oda}}, \bibinfo {author} {\bibfnamefont {O.}~\bibnamefont {Ivashko}}, \bibinfo {author} {\bibfnamefont {M.~v.}\ \bibnamefont {Zimmermann}}, \bibinfo {author} {\bibfnamefont {M.}~\bibnamefont {Janoschek}},\ and\ \bibinfo {author} {\bibfnamefont
  {J.}~\bibnamefont {Chang}},\ }\bibfield  {title} {\bibinfo {title} {Unveiling unequivocal charge stripe order in a prototypical cuprate superconductor},\ }\href {https://doi.org/10.1103/PhysRevLett.128.207002} {\bibfield  {journal} {\bibinfo  {journal} {Phys. Rev. Lett.}\ }\textbf {\bibinfo {volume} {128}},\ \bibinfo {pages} {207002} (\bibinfo {year} {2022})}\BibitemShut {NoStop}%
\bibitem [{\citenamefont {Wang}\ \emph {et~al.}(2022)\citenamefont {Wang}, \citenamefont {von Arx}, \citenamefont {Mazzone}, \citenamefont {Mustafi}, \citenamefont {Horio}, \citenamefont {K{\"u}spert}, \citenamefont {Choi}, \citenamefont {Bucher}, \citenamefont {Wo}, \citenamefont {Zhao}, \citenamefont {Zhang}, \citenamefont {Asmara}, \citenamefont {Sassa}, \citenamefont {M{\aa}nsson}, \citenamefont {Christensen}, \citenamefont {Janoschek}, \citenamefont {Kurosawa}, \citenamefont {Momono}, \citenamefont {Oda}, \citenamefont {Fischer}, \citenamefont {Schmitt},\ and\ \citenamefont {Chang}}]{Wang2022}%
  \BibitemOpen
  \bibfield  {author} {\bibinfo {author} {\bibfnamefont {Q.}~\bibnamefont {Wang}}, \bibinfo {author} {\bibfnamefont {K.}~\bibnamefont {von Arx}}, \bibinfo {author} {\bibfnamefont {D.~G.}\ \bibnamefont {Mazzone}}, \bibinfo {author} {\bibfnamefont {S.}~\bibnamefont {Mustafi}}, \bibinfo {author} {\bibfnamefont {M.}~\bibnamefont {Horio}}, \bibinfo {author} {\bibfnamefont {J.}~\bibnamefont {K{\"u}spert}}, \bibinfo {author} {\bibfnamefont {J.}~\bibnamefont {Choi}}, \bibinfo {author} {\bibfnamefont {D.}~\bibnamefont {Bucher}}, \bibinfo {author} {\bibfnamefont {H.}~\bibnamefont {Wo}}, \bibinfo {author} {\bibfnamefont {J.}~\bibnamefont {Zhao}}, \bibinfo {author} {\bibfnamefont {W.}~\bibnamefont {Zhang}}, \bibinfo {author} {\bibfnamefont {T.~C.}\ \bibnamefont {Asmara}}, \bibinfo {author} {\bibfnamefont {Y.}~\bibnamefont {Sassa}}, \bibinfo {author} {\bibfnamefont {M.}~\bibnamefont {M{\aa}nsson}}, \bibinfo {author} {\bibfnamefont {N.~B.}\ \bibnamefont {Christensen}}, \bibinfo {author} {\bibfnamefont {M.}~\bibnamefont
  {Janoschek}}, \bibinfo {author} {\bibfnamefont {T.}~\bibnamefont {Kurosawa}}, \bibinfo {author} {\bibfnamefont {N.}~\bibnamefont {Momono}}, \bibinfo {author} {\bibfnamefont {M.}~\bibnamefont {Oda}}, \bibinfo {author} {\bibfnamefont {M.~H.}\ \bibnamefont {Fischer}}, \bibinfo {author} {\bibfnamefont {T.}~\bibnamefont {Schmitt}},\ and\ \bibinfo {author} {\bibfnamefont {J.}~\bibnamefont {Chang}},\ }\bibfield  {title} {\bibinfo {title} {Uniaxial pressure induced stripe order rotation in {La$_{1.88}$Sr$_{0.12}$CuO$_4$}},\ }\href {https://doi.org/10.1038/s41467-022-29465-4} {\bibfield  {journal} {\bibinfo  {journal} {Nature Communications}\ }\textbf {\bibinfo {volume} {13}},\ \bibinfo {pages} {1795} (\bibinfo {year} {2022})}\BibitemShut {NoStop}%
\bibitem [{\citenamefont {Gupta}\ \emph {et~al.}(2023)\citenamefont {Gupta}, \citenamefont {Sutarto}, \citenamefont {Gong}, \citenamefont {Idziak}, \citenamefont {Hale}, \citenamefont {Kim},\ and\ \citenamefont {Hawthorn}}]{gupta2023}%
  \BibitemOpen
  \bibfield  {author} {\bibinfo {author} {\bibfnamefont {N.~K.}\ \bibnamefont {Gupta}}, \bibinfo {author} {\bibfnamefont {R.}~\bibnamefont {Sutarto}}, \bibinfo {author} {\bibfnamefont {R.}~\bibnamefont {Gong}}, \bibinfo {author} {\bibfnamefont {S.}~\bibnamefont {Idziak}}, \bibinfo {author} {\bibfnamefont {H.}~\bibnamefont {Hale}}, \bibinfo {author} {\bibfnamefont {Y.-J.}\ \bibnamefont {Kim}},\ and\ \bibinfo {author} {\bibfnamefont {D.~G.}\ \bibnamefont {Hawthorn}},\ }\href@noop {} {\bibinfo {title} {Tuning charge density wave order and structure via uniaxial stress in a stripe-ordered cuprate superconductor}} (\bibinfo {year} {2023}),\ \Eprint {https://arxiv.org/abs/2305.16499} {arXiv:2305.16499 [cond-mat.str-el]} \BibitemShut {NoStop}%
\end{thebibliography}
%
\end{document}